\documentclass[journal]{IEEEtran}
\usepackage[latin1]{inputenc}
\usepackage{times,amsmath}
\usepackage{comment}
\usepackage{amssymb}
\usepackage{steinmetz}
\usepackage{pstool}
\usepackage{multirow}
\usepackage{enumerate}
\usepackage{graphicx}
\usepackage{multirow}
\usepackage{booktabs}
\usepackage{tabularx}
\usepackage[table]{xcolor}
\usepackage{subcaption}
\usepackage{mwe}
\usepackage{bigints}
\usepackage{MnSymbol}
\usepackage{stfloats} 
\usepackage[square, comma, sort&compress, numbers]{natbib}
\usepackage{nohyperref}
\usepackage{algorithm,algorithmic}
\usepackage{epstopdf}
\usepackage{mathtools, cuted}

\usepackage[square, comma, sort&compress, numbers]{natbib}
\usepackage{amsfonts}
\usepackage{multicol}
\usepackage{diagbox}
\usepackage{array}

\graphicspath{ {Figures/} }

\definecolor{light-gray}{gray}{0.85}
\definecolor{c1}{RGB}{240,248,255}
\definecolor{c2}{RGB}{245,245,220}
\definecolor{c3}{RGB}{255,228,225}

\begin{document}

\title{Radio-PPG: photoplethysmogram digital twin synthesis using deep neural representation of 6G/WiFi ISAC signals}

\author{
\IEEEauthorblockN{
Israel Jesus Santos Filho\IEEEauthorrefmark{1}, Muhammad Mahboob Ur Rahman\IEEEauthorrefmark{1}, Taous-Meriem Laleg-Kirati\IEEEauthorrefmark{2}, Tareq Al-Naffouri\IEEEauthorrefmark{1} }

\IEEEauthorblockA{\IEEEauthorrefmark{1}Computer, Electrical and Mathematical Sciences and Engineering Division (CEMSE), \\ 
King Abdullah University of Science and Technology, Thuwal 23955, Saudi Arabia.\\ 
\IEEEauthorrefmark{2}The National Institute for Research in Digital Science and Technology, Paris-Saclay, France.\\
\IEEEauthorrefmark{1}\{israel.filho,muhammad.rahman,tareq.alnaffouri\}@kaust.edu.sa, \IEEEauthorrefmark{2}Taous-Meriem.Laleg@inria.fr }
\thanks{
The research reported in this publication was supported by funding from King Abdullah University of Science and Technology (KAUST) - KAUST Center of Excellence for Smart Health (KCSH), under award number 5932.
} 
}


\maketitle

\begin{abstract} 



Digital twins for 1D bio-signals enable real-time monitoring of physiological processes of a person, which enables early disease diagnosis and personalized treatment. This work introduces a novel non-contact method for digital twin (DT) photoplethysmogram (PPG) signal synthesis under the umbrella of 6G/WiFi integrated sensing and communication (ISAC) systems. We employ a software-defined radio (SDR) operating at 5.23 GHz that illuminates the chest of a nearby person with a wideband 6G/WiFi signal and collects the reflected signals. This allows us to acquire Radio-PPG dataset that consists of 300 minutes worth of near synchronous 64-channel radio data, PPG data, along with the labels (three body vitals) of 30 healthy subjects. With this, we test two artificial intelligence (AI) models for DT-PPG signal synthesis: i) discrete cosine transform followed by a multi-layer perceptron, ii) two U-NET models (Approximation network, Refinement network) in cascade, along with a custom loss function. Experimental results indicate that U-NET model achieves an impressive relative mean absolute error of 0.194 with a small ISAC sensing overhead of 15.62\%, for DT-PPG synthesis. Furthermore, we performed quality assessment of the synthetic DT-PPG by computing the accuracy of DT-PPG-based vitals estimation and feature extraction, which turned out to be at par with that of reference PPG-based vitals estimation and feature extraction.
This work highlights the potential of generative AI and 6G/WiFi ISAC technologies and serves as a foundational step towards the development of non-contact screening tools for covid-19, cardiovascular diseases and well-being assessment of people with special needs.

\end{abstract}


\section{Introduction}
\label{sec:intro}

Cardiovascular health digital twin (CHDT) is an emerging paradigm that has gained lots of attention recently, thanks to the massive advancements in the fields of artificial intelligence (AI), internet of medical things (IoMT) sensing, and edge/cloud computing \cite{coorey2022health}. A CHDT system consists of a sensing block (basically, a suite of IoMT sensors, e.g., wearables, implants, etc., which do various kinds of cardiovascular health sensing), an inference block (sitting at the cloud that conditions and fuses the sensed data through various signal processing and artificial intelligence methods), and a display panel (a smartphone app, or a computer dashboard that displays the inference results) \cite{martinez2019cardio}. Thus, the CHDT systems aim to create a detailed digital replica of cardiovascular system of a person, by means of synthesis of either simple 1D biosignals or sophisticated 2D images \cite{chakshu2021towards,viola2023gpu}. CHDT systems, due to their continuous monitoring nature, are capable of capturing the fine-grained anatomical and physiological changes in the cardiovascular system of a person over time \cite{corral2020digital}. This enables the CHDT systems to offer a number of advantages, e.g., early prediction of a cardiovascular disease (CVD) such as arrhythmia, myocardial infarction \cite{li2024towards}, chronic heart disease management, clinical decision support for doctors, monitoring of recovery rate after heart surgery, precision medicine \cite{coorey2021health}, to name a few. Despite the challenges (such as privacy issues, government regulations), CHDT systems have the potential to revolutionize cardiovascular healthcare, offering more personalized, predictive, and preventive care.  

Inline with previous work \cite{mazumder2019synthetic}, this work aims to create a digital twin representation of photoplethysmography (PPG) signal, which is a biomarker of great clinical significance. 
PPG is traditionally acquired in a non-invasive manner from the fingertip or earlobe of a person by means of a pulse oximeter which utilizes optical principles to measure changes in flow of blood volume in peripheral veins \cite{ppg_soa_2021}. PPG is frequently used in various settings, from healthcare facilities to wearable devices, to carry out a wide range of health analytics \cite{almarshad2022diagnostic}, e.g., estimation of body vitals (heart rate, blood Oxygen saturation (SpO2), respiratory rate) \cite{mehmood2023your}, blood pressure estimation \cite{wang2018towards}, sleep quality, stage, apnea, insomnia analysis \cite{vulcan2021photoplethysmography}, CVD diagnosis \cite{ouyang2021use}, dehydration monitoring \cite{alaslani2024you}, vascular age estimation \cite{khalid2023low}, arterial stiffness measurement, and more. 

The non-invasive nature of PPG allows frequent, in-situ measurements (by means of oximeters and cameras). This is what has motivated us to create a CHDT representation of PPG for real-time monitoring of cardiovascular health. 
However, this work takes a step forward as it synthesizes the PPG signal of a person in a contactless manner through 5G/6G/WiFi signals, capitalizing on integrated sensing and communication framework (ISAC) of 6G cellular systems \cite{kaushik2024integrated}. The motivation for this comes from the fact that contact-based acquisition of PPG might not be possible under certain circumstances, e.g., health monitoring of certain vulnerable population groups (e.g., newborns, Autistic individuals, patients in intensive care units, elderly), during the outbreak of pandemics, e.g., covid19 \cite{abu2024contactless,chen2024monitoring}. In such situations, it becomes imperative to develop a new class of CHDT systems that are empowered through generative AI and 6G-based non-contact ISAC methods, and could measure physiological phenomenon of interest from a distance. Non-contact CHDT systems are anticipated to be an integral part of smart homes, and smart cities of future \cite{umair2021impact}.



\subsection{Related Work}

Broadly speaking, related work consists of two categories: i) the works that design non-contact health sensing methods, and ii) the works that aim to do biomedical waveform translation, both under the umbrella of CHDT. Below, we discuss selected related work from both categories.

{\it Non-contact health sensing methods:}
Non-contact health sensing (NCHS) methods have seen a sharp rise in popularity in the post-covid19 era due to the fact that they are inline with the community guidelines of maintaining a safe distance \cite{taylor2020review}. NCHS methods utilize a number of wireless sensing modalities, e.g., software-defined radios (SDR) \cite{lopes2024covis}, frequency-modulated continuous-wave radars and ultra-wideband radars \cite{paterniani2023radar, ahmed2023machine}, security cameras and regular cameras \cite{romano2021non, paul2020non, kumar2015distanceppg}, WiFi signals \cite{ge2022contactless}, radio-frequency identification tags \cite{khan2024tag}, etc., to solve a wide range of contactless health sensing problems\footnote{Researchers have investigated other contactless health sensing modalities as well, e.g., ultrasound-based methods \cite{ambrosanio2019multi}. But this work focuses on radio signals-based methods only.}. For example, researchers have done vitals estimation \cite{romano2021non,paterniani2023radar,ahmed2023machine,massaroni2020contactless}, sleep quality analysis \cite{hsu2017zero}, fall detection \cite{khan2024tag}, gesture recognition \cite{liu2021m}, posture tracking \cite{yue2020bodycompass}, gait analysis \cite{di2022markerless}, dehydration monitoring \cite{buttar2023non}, breathing abnormalities detection \cite{pervez2023hand}, lung disease classification \cite{buttar2024non}, covid19 related surveillance and analytics \cite{murad2022wireless}, using the aforementioned radio sensing modalities. Furthermore, there are works that provide an extensive discussion of the existing public datasets for NCHS methods \cite{liang2024review}. 

{\it Biomedical waveform reconstruction methods:}
The recent rise in generative AI methods has prompted the researchers to reconstruct a number of clinically-significant biomedical signals under the umbrella of CHDT. Specifically, there are works that aim to do PPG to electrocardiogram (ECG) translation \cite{vo2021p2e,guo2024unet,zhu2021learning}, PPG to arterial blood pressure (ABP) translation \cite{mehrabadi2022novel,ma2023ppg,tahir2024cuff}, phonocardiogram (PCG) to ECG translation \cite{dissanayake2022generalized}, radar signal to ECG translation \cite{chowdhury2024ecg,9919401}, radar signal to seismocardiogram (SCG) translation \cite{ha2020contactless}, radar signal to PPG translation \cite{khan2022contactless}. In addition, there are works that denoise and reconstruct PPG and other biomedical signals using generative adversarial networks (GAN) \cite{wang2022ppg, long2023reconstruction}, using hybrid wavelet-deep learning methods \cite{ahmed2023deep}.


{\it However, to the best of authors' knowledge, non-contact acquisition of digital twin PPG based on 6G/WiFi/SDR sensing has not been done before.}

\subsection{Contributions}

This work is the first to demonstrate the feasibility of utilizing the existing WiFi and cellular (4G/5G) infrastructure to synthesize the digital twin PPG of a person in a contactless manner through cellular/WiFi signals. Specifically, we utilize an SDR-based system that radio-exposes a person, collects the reflected radio signal, and utilizes the generative AI tools to synthesize the digital twin PPG signal of that person. 

{\it Feasibility:}
The problem at hand, i.e., digital twin PPG synthesis through 6G/WiFi signals, is indeed a feasible problem. This is due to the fact that the 6G/WiFi signals utilize microwave frequencies, which help them penetrate deep inside the human chest; this in turn allows them to faithfully capture the rhythmic movement of the heart and the lungs. Secondly, the 6G/WiFi signals are broadband, and thus, are well-suited for the ISAC framework of future 6G systems whereby the communication signals are to be used for health sensing. Specifically, the orthogonal frequency division multiplexing (OFDM) signals consists of a number of channels which together observe the human chest movement at many different frequencies, and thus, help synthesize the digital twin PPG. 

The key contributions of this work are as follows:
\begin{itemize}
    \item {\it Radio-PPG dataset:} This work presents Radio-PPG dataset, first of its kind, that provides nearly synchronous recordings of the 6G/WiFi OFDM signals (at 5.23 GHz) and red-channel PPG signals\footnote{A preliminary version of this research was presented at the IFAC BMS 2024 conference \cite{filho2024noncontact}.}. More precisely, the dataset consists of recordings of raw 64-channel OFDM and red-channel PPG signal (the ground truth), acquired from 30 young subjects (15 males, 15 females). Further, we simultaneously acquired the body vitals (heart rate, SpO2, breathing rate) of the subjects through a pulse oximeter, with the purpose of subsequent validation of the synthetic digital twin PPG signal. 
    \item {\it Digital twin PPG synthesis from 6G/WiFi signals:}
    A custom-built signal processing pipeline is utilized to pre-process the raw 6G/WiFi OFDM data, i.e., channel frequency response. Afterwards, the conditioned data is fed to two custom deep learning (DL) models, which perform digital twin PPG synthesis as follows: 1) The first approach learns a non-linear mapping between the conditioned OFDM signal and reference PPG signal in the frequency domain, through discrete cosine transform. 2) The second approach utilizes a custom-built U-NET convolutional model that fuses OFDM signals (across 64 channels) in the time domain, and produces digital twin PPG waveform through a two-step process involving an approximation network and a refinement network, both based on U-NET structure.
    \item {\it Validation of the digital twin PPG:} The fidelity of the synthetic digital twin PPG is validated by means of two experiments as follows: i) we estimate body vitals (heart rate, SpO2 and respiratory rate) from both the digital twin PPG and the reference PPG, using another custom-built DL model, ii) we do feature extraction (from PPG and its four derivative signals) from both the digital twin PPG and the reference PPG. For both experiments, synthetic digital twin PPG performs at par with the reference PPG. 
\end{itemize}

The proposed method also has the potential to be deployed at scale. That is, it could allow the competent authorities to create, monitor and track digital twin PPG representation of a target population, e.g., a hotspot neighborhood during pandemic outbreaks, etc. 




\subsection{Outline} 

{The rest of this paper is organized as follows: Section II details the experimental setup used to acquire the Radio-PPG dataset. Section III describes in detail the pre-processing pipeline that we have developed to condition the OFDM and PPG signals. Section IV discusses the deep learning architectures that we developed for digital twin PPG waveform synthesis. Section V provides detailed performance analysis. Section V concludes the paper. 
}

\section{Acquisition of Radio-PPG dataset}
\label{sec:method}


Note that there exists no publicly available dataset that provides synchronous measurements of radio signals and PPG signals, which poses a major barrier to reproducibility and benchmarking in this domain (see Table \ref{tab:dataset_comparison}). This research gap motivated us to construct a custom dataset, specifically designed to enable the development of a PPG digital twin from 6G/WiFi-like signals. To the best of our knowledge, this is the first dataset that facilitates the construction of a PPG digital twin, whereas existing RF/WiFi sensing datasets focus primarily on coarse vital sign monitoring (e.g., respiration, heart rate) and remain private. This also implies that direct comparisons with prior work are not fully meaningful, since existing studies differ significantly in data modality, subject demographics, and collection environments. Our dataset therefore helps us evaluate the performance of the proposed PPG digital twin methods in this work, and is a step toward advancing generative AI research in the broad domain of WiFi-based health sensing. Furthermore, the experimental setup used in this work for data collection is cost-effective, and could be deployed at scale.

For data acquisition, we set up a universal software radio peripheral (USRP) SDRs-based 6G ISAC OFDM link. Specifically, we utilized two USRP N210 SDRs, each connected to a PC and a directional horn antenna, in order to establish a 64-channel OFDM link of bandwidth 200 KHz at a center frequency of 4 GHz. It is worth mentioning that OFDM is the default multi-channel modulation technique in 5G/6G/WiFi systems, whereby the data is transmitted over multiple orthogonal channels. The succinct details about theoretical foundations of a 6G/WiFi OFDM link, as well as the key hyper-parameters of the OFDM link deployed in this work are provided in Appendix A.

\begin{figure}[ht]
\begin{center}
	\includegraphics[width=0.9\linewidth]{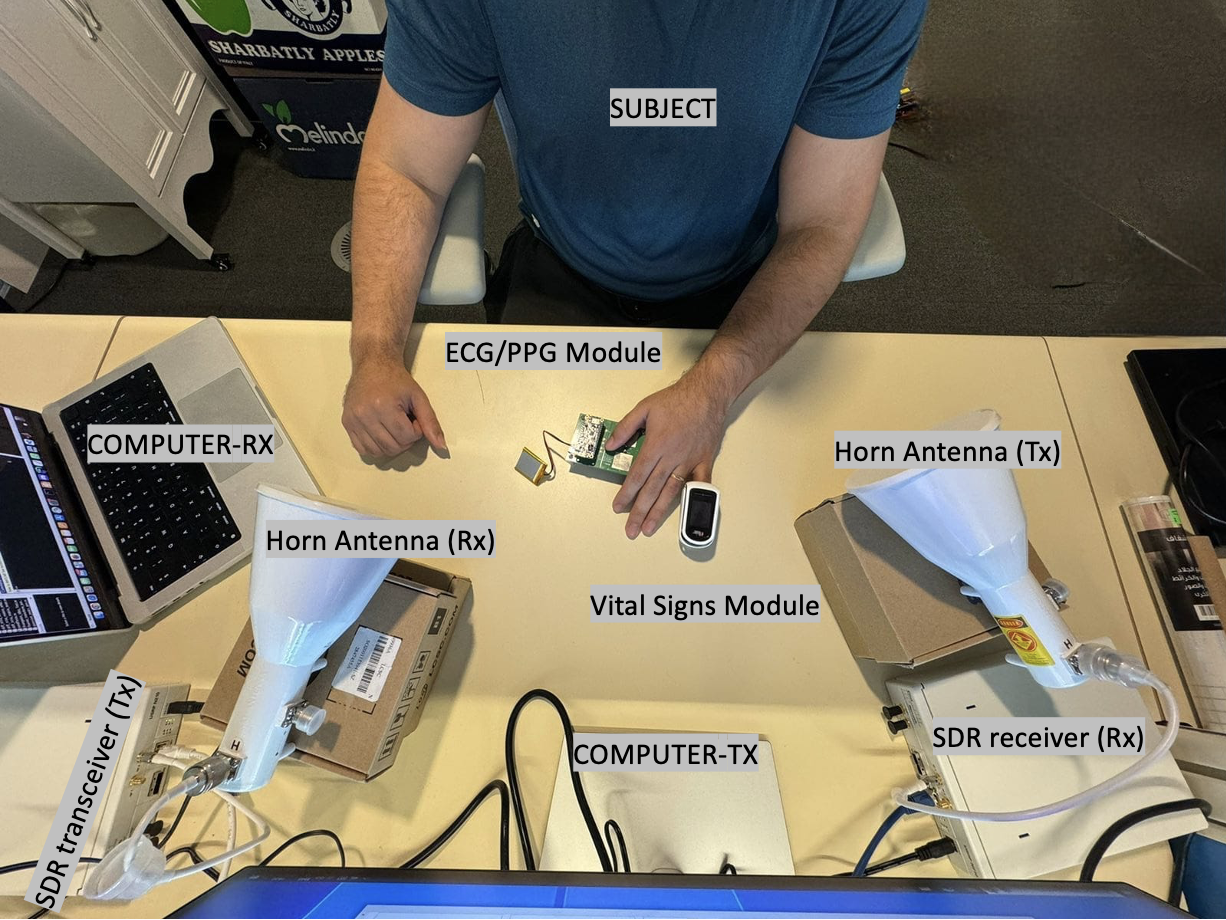} 
\caption{Experimental setup for near-simultaneous collection of the raw 64-channel OFDM CFR data (through 6G/WiFi link), reference PPG data (through MAX86150 module), and body vitals (through Massimo pulse oximeter). }
\label{fig:instrument_collection}
\end{center}
\end{figure}

Once the OFDM link was operational, we started the data acquisition process whereby we radio-exposed (by means of the OFDM transmission) the chest area of each subject who sat close to the table that hosted the OFDM transmitter and receiver (see Fig. \ref{fig:instrument_collection}). One horn antenna directed the tramsmit beam towards the subject, while the other horn antenna collected the OFDM signal reflected off the chest of the subject. Throughout the duration of experiment, we instructed the subjects to sit still in order to avoid motion-induced artefacts in the data being gathered. In addition to collection of 64-channel OFDM signals (the raw data) through horn antenna, we also acquired the single-channel PPG signal (the reference waveform) through MAX86150 module, and the body vitals (i.e., heart rate, SpO2, respiratory rate) through Massimo pulse oximeter, in a controlled manner, in order to ensure synchronization. Note that the 64-channel OFDM data, single-channel reference PPG signal, body vitals were collected at a sampling rate of 20 KHz, 200 Hz, 1 Hz, respectively. Further, the MAX86150 module (pulse oximeter) was connected via Bluetooth to a PC (smartphone) for logging of reference PPG waveform (body vitals). 

Finally, we present a quick summary of the key statistics of the dataset that we have constructed.
The custom Radio-PPG dataset consists of labeled data (64-channel OFDM data, reference single-channel PPG data, vitals data) from 30 healthy and young volunteers, 15 males (aged 22-36 years), and 15 females (aged 22-32 years). For each subject, we collected data twice, each of duration 5 minutes, leading to a total of 300 minutes of labeled data.\footnote{This research study was approved by the Institutional Biosafety and Bioethics Committee (IBEC) of King Abdullah University of Science and Technology, Saudi Arabia, Protocol number: 23IBEC002, date of approval: Sept. 27, 2023. All subjects provided their written informed consent before the data collection. The data collection was conducted in accordance with the Declaration of Helsinki.}

\begin{table*}[h!]
\centering
\caption{Comparative Analysis of Datasets for Radio/WiFi-Based Physiological Sensing}
\label{tab:dataset_comparison}
\resizebox{\textwidth}{!}{%
\begin{tabular}{l l l l c p{5cm}}
\toprule
\textbf{Work / Dataset Title} & \textbf{Modality} & \textbf{Dataset Size (hours)} & \textbf{Frequency Spectrum} & \textbf{Participants} & \textbf{Physiological Signal(s) Reproduced} \\
\midrule
\textbf{Radio-PPG (this work)} & \textbf{OFDM} & \textbf{5} & \textbf{5.24 GHz (Wi-Fi OFDM)} & \textbf{30} & \textbf{PPG Waveform, Vitals} \\
Non-Contact PPG (Filho et al., 2024) & OFDM & 2.67 & 5.24 GHz (Wi-Fi OFDM) & 16 & PPG Waveform \cite{filho2024noncontact} \\
WiFi Sleep Monitoring (Ali et al., 2021) & Wi-Fi CSI & $>$550 & 2.4/5 GHz & 5 & Respiration Rate, Body Motion \cite{ali2021goodness} \\
PhaseBeat (Wang et al., 2017) & Wi-Fi CSI & Not Specified & 5 GHz Band & 4 & Respiration Rate, Heart Rate \cite{wang2017phasebeat} \\
MMECG (Chen et al., 2022) & mmWave Radar & 10 & 77-81 GHz (FMCW) & 35 & ECG Waveform \cite{chen2022contactless} \\
PhysDrive (Wang et al., 2025) & mmWave Radar & $>$41 (Est.) & 77-81 GHz (FMCW) & 48 & ECG, BVP Waveforms, Respiration \cite{wang2025physdrive} \\
FMCW Radar (Wang et al., 2024) & mmWave Radar & Not Specified & Not Specified & Not Specified & Heart Rate, Breathing Rate \cite{wang2024novel} \\
IR-UWB Radar (Lee et al., 2023) & UWB Radar & Not Specified & IR-UWB & N/A & Heart Rate, Respiration Rate \cite{lee2023deep} \\
\bottomrule
\end{tabular}%
}
\end{table*}


\section{{Data Pre-processing}}

The primary objective of this work is to address a fundamentally novel signal translation problem: synthesizing a physiological PPG waveform directly from a WiFi signal. This task is distinct from traditional PPG analysis, which typically focuses on mitigating motion artifacts within the noisy PPG signal. Here, the core challenge lies in extracting faint modulations caused by physiological processes from a complex, multi-channel radio signal that is corrupted by environmental reflections. Therefore, a pre-processing pipeline is necessary not only for the denoising step but also as a critical component designed to isolate and condition the relevant radio signal components for the subsequent PPG synthesis task. This section details the framework developed to condition both the raw OFDM data and the reference PPG signal, ensuring they are suitable for a cross-modal translation.

\subsubsection{{Pre-processing of OFDM signal}}

To isolate the embedded physiological information from the raw radio data, we perform a series of targeted pre-processing steps as follows. In our previous work \cite{filho2024noncontact}, we applied a traditional signal processing pipelines that fused multi-channel data into a single waveform \textit{prior} to modeling. Now, our new approach aligns with modern deep learning practices where reconstruction is guided by deep feature mapping. We preserve the rich, multi-channel information from the OFDM signal, allowing the neural network to learn the underlying features and their complex correlations directly. This strategy is enforced by our pre-processing, which focuses on structuring the data rather than aggressive, manual feature extraction. The pipeline is as follows:

\begin{itemize}
 \item \textbf{Channel Selection:} {The raw OFDM signal consists of 64 complex-valued channels\footnote{Note that the terms radio data, OFDM data, and CFR data all refer to the raw 6G/WiFi data collected through USRP N210 SDRs; therefore, we use these terms interchangeably throughout the rest of this paper.}. As adjacent channels provide highly correlated information, we first perform a preliminary dimensionality reduction by selecting 16 representative channels out of the 64 available, reducing redundancy while retaining frequency diversity.}
 
 \item \textbf{Channel Expansion:} {The 16 selected OFDM channels are complex-valued. To leverage the complete signal information for the deep learning model, we separate each channel into its real and imaginary components. This expansion transforms the 16 complex channels into a 32-channel real-valued tensor (16 real parts and 16 imaginary parts). This crucial step ensures that both phase and magnitude information are preserved, providing a much richer input for the subsequent neural network to guide the point-to-point waveform reconstruction.}
  
 \item \textbf{Data Segmentation:} {The 32-channel radio data is then segmented into non-overlapping windows of 2.5 seconds. Each segment now represents a multi-channel snapshot of the chest's movement, ready for batch processing.}
 
 \item \textbf{Data Normalization:} {We then normalize each of the 32 channels independently across the time dimension using the z-score method. This ensures that all input channels have a consistent scale without distorting their temporal dynamics.}
 
 \item \textbf{Data Augmentation:} {Finally, to enhance the robustness and generalizability of our models, we augment the data by a factor of 2. This is achieved by adding Gaussian noise (mean 0, variance 0.01) to the original 32-channel segments, creating noisy copies.}

\end{itemize}

\begin{figure}[ht]
\centering
	\includegraphics[width=10cm,height=5cm]{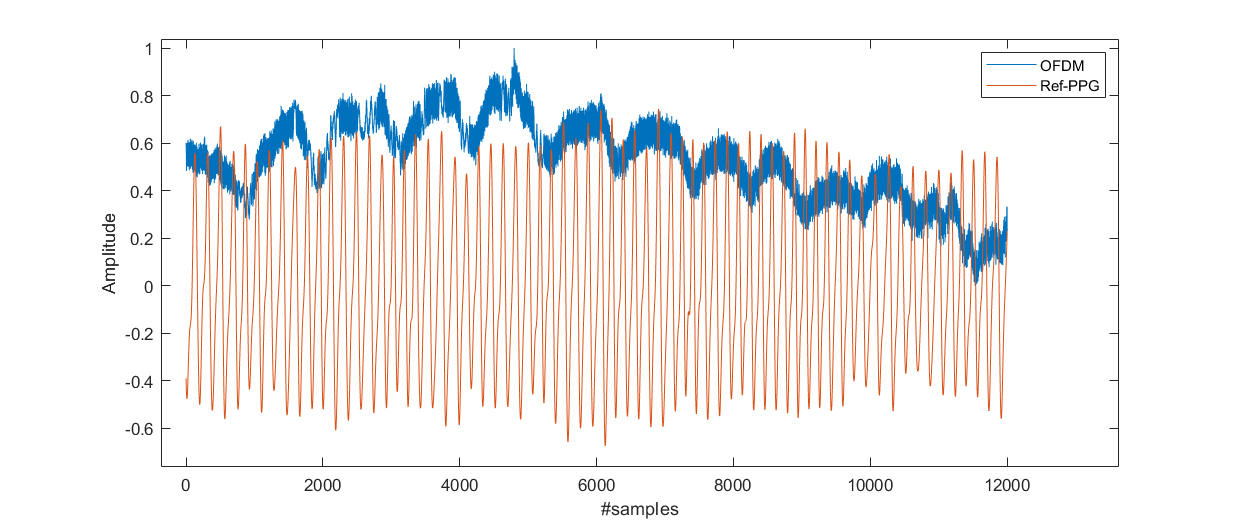} 
\caption{The magnitude signal of a single OFDM channel superimposed on the reference PPG.}
\label{fig:ppg_ofdm_after_pca}
\end{figure}

\begin{figure}[ht]
\centering
	\includegraphics[width=10cm, height=5.3cm]{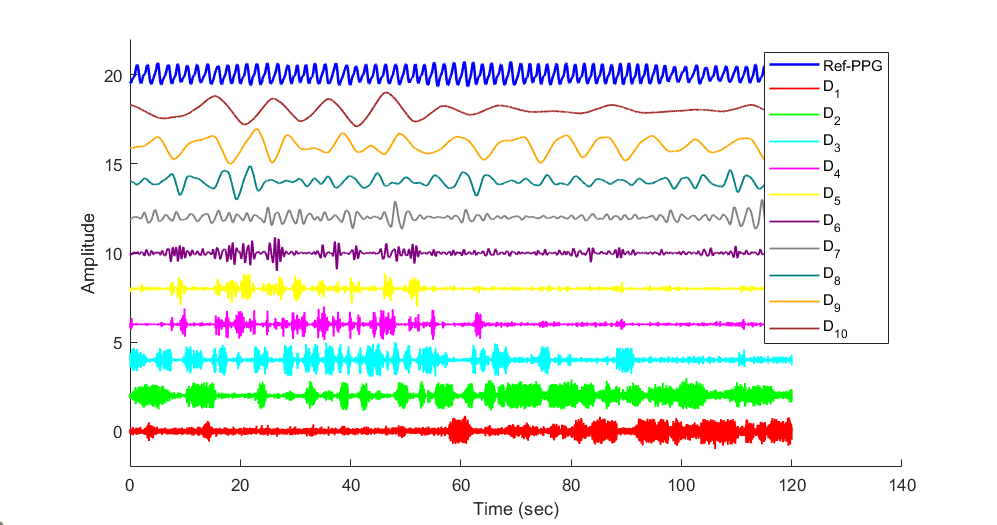}
\caption{The discrete wavelet decomposition block returns the wavelet component $D_7$ that is frequency-matched with the reference PPG signal.}
\label{fig:D7_wavelet}
\end{figure}

\subsubsection{{Pre-processing of reference PPG signal}}

The reference PPG signal, though from a direct optical source, is also susceptible to noise (e.g., baseline wander) and artifacts (e.g., from motion). We apply a standard conditioning pipeline to ensure a clean ground-truth signal.
\begin{enumerate}
 \item \textbf{Detrending:} We remove baseline drift by estimating the low-frequency trend using a wavelet transform (specifically, the \textit{db2} wavelet family) and subtracting it from the original signal.
 
 \item \textbf{Denoising:} Next, we apply a 12th-order lowpass Butterworth filter with a 4 Hz cut-off to remove out-of-band noise.
 
 \item \textbf{Data segmentation:} We segment the PPG data into 2.5-second windows to correspond with the radio data segments.
 
 \item \textbf{Data normalization:} Each segment is then normalized using the z-score method.
 
 \item \textbf{Data augmentation:} Finally, we augment the PPG data identically to the radio data, adding Gaussian noise to create a corresponding set of noisy copies.
\end{enumerate}

\subsubsection{Alignment and Synchronization}

\begin{figure}[ht]
\centering
	\includegraphics[width=9cm, height=7cm]{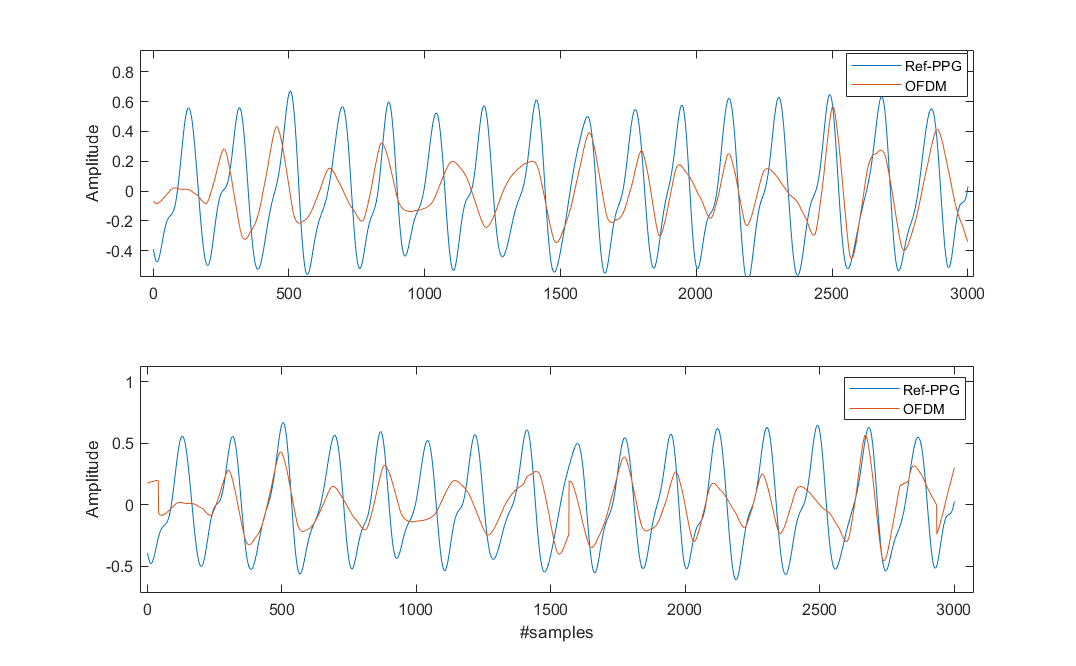}
\caption{Illustration of the waveform alignment process. Left: The signals before alignment show a clear phase offset. Right: After compensating for the calculated time lag $\tau$, the signals are perfectly synchronized.}
\label{fig:alignment}
\end{figure}

A critical challenge in cross-modal analysis is ensuring precise temporal alignment. The two conditioned signals, i.e., the $D_7$ wavelet component from the OFDM signal and the reference PPG, exhibit a slight time offset due to minor delays during data acquisition (see Fig. \ref{fig:alignment}, left). This phase mismatch, if uncorrected, would severely impair the performance of any translation model.

To compensate for this, we formulate an optimization problem to maximize the alignment between the two signals. We find the time lag $\tau$ that maximizes the inner product between the radio-derived waveform $w(t)$ (channel-wise) and the time-shifted reference PPG $y(t-\tau)$:
\begin{equation}
\label{aling_optimization}
\max_{\tau} \langle y(t-\tau), w(t) \rangle = \max_{\tau} \int y(t-\tau) w(t) \, dt,
\end{equation}
where $\langle \cdot, \cdot \rangle$ denotes the inner product. Solving this template matching problem provides the exact time lag, which we then use to perfectly synchronize the two signals, as shown in Fig. \ref{fig:alignment} (bottom). This final step yields a dataset of partially aligned signal pairs.

\section{{Synthesis of Digital Twin PPG Waveform}}

\begin{figure*}[ht]
\begin{center}
	\includegraphics[width=16cm,height=9cm]{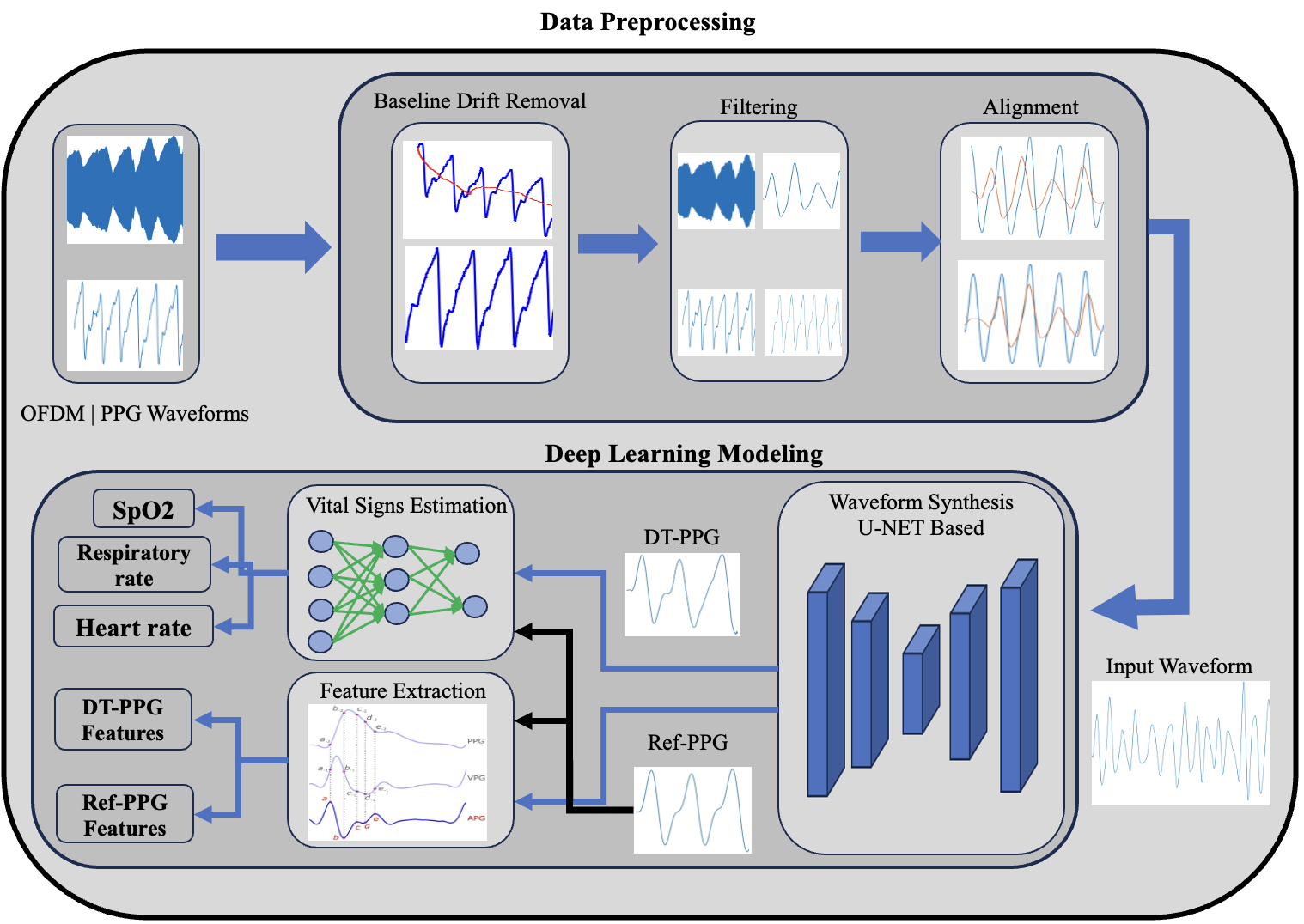} 
\caption{Proposed hybrid signal processing+AI method for digital twin PPG synthesis from 6G/WiFi OFDM signals.}
\label{fig:complete_pipalig_sig}
\end{center}
\end{figure*}

The digital twin PPG synthesis problem aims to learn a map between the pre-processed OFDM waveform and the reference PPG waveform. Let $x(t)$ represent the OFDM waveform and let $y(t)$ represent the reference PPG waveform, after pre-processing. Further, let $\mathcal{T}_{\theta}$ represent the transformation operator, basically a neural network with parameters $\theta$ (i.e., the weights and biases of the neurons in the neural network). Then, the following holds: $y(t) = \mathcal{T}_\theta(x(t))$. 
To this end, we learn the non-linear transformation operator $\mathcal{T}_{\theta}$ between the OFDM waveform and the reference PPG waveform using two distinct approaches: 1) discrete cosine transform (DCT) and multi layer perceptron (MLP)-based frequency-domain (FD) approach. 2) U-NET model-based time-domain (TD) approach. 
Fig. \ref{fig:complete_pipalig_sig} presents the complete block diagram of the proposed hybrid method that consists of a signal pre-processing pipeline, followed by two deep learning frameworks for digital twin PPG synthesis, followed by the validation stage. 

\subsection{{Baseline approach: DCT + MLP-based FD method}}

Our first approach serves as a baseline, learning the signal mapping in the frequency domain. It begins by transforming both the OFDM segment and the reference PPG segment to the frequency domain using DCT type-II:
\begin{equation} \label{dct_revised}
 X[k] = \sum_{n=0}^{N-1} x[n] \cos \left( \frac{\pi}{N} \left( n + \frac{1}{2} \right) k \right) \quad \text{for } n,k = 0, 1, \ldots, N-1
\end{equation}
where $x[n]$ is the signal of interest, and $X[k]$ is the vector of DCT coefficients.

{Next, we learn the non-linear map between the two sets of DCT coefficients using a five-layer MLP, with architectural details summarized in Table \ref{tab:mlp}. The model utilizes the {Gaussian Error Linear Unit (GELU)} activation function \cite{hendrycks2016gaussian}. Unlike ReLU, which deterministically gates neurons, GELU weights inputs based on their magnitude, creating a smoother, probabilistic activation curve that has shown strong performance in modern deep learning models. Mathematically, GELU is defined as}:
\begin{equation}
\label{eq:gelu_exact}
\text{GELU}(x) = x \cdot \Phi(x) = \frac{1}{2} x \left(1 + \text{erf}\left(\frac{x}{\sqrt{2}}\right)\right)
\end{equation}
{where $\Phi(x)$ is the cumulative distribution function (CDF) of the standard normal distribution, and $\text{erf}$ is the error function. In practice, a fast approximation is often used to improve computational performance}:
\begin{equation}
\label{eq:gelu_approx}
\text{GELU}(x) \approx 0.5x \left(1 + \tanh\left[\sqrt{2/\pi}(x + 0.044715x^3)\right]\right)
\end{equation}

\begin{table}[h!]
\centering
\begin{tabular}{|c|c|c|c|}
\hline
\textbf{Layer} & \textbf{Type} & \textbf{Output Shape} & \textbf{Activation/Dropout} \\ 
\hline
1  & \textbf{Input}& ($b_s, N_{ch} \times N \times K$) & None \\
\hline
2  & \textbf{Linear}  & ($b_s, \text{2048}$)& None \\
\hline
3  & \textbf{Dropout} & ($b_s, \text{2048}$)& Dropout (0.05) \\
\hline
4  & \textbf{Activation} & ($b_s, \text{2048}$)& GELU\\
\hline
5  & \textbf{Linear}  & ($b_s, \text{1024}$)& None \\
\hline
6  & \textbf{Dropout} & ($b_s, \text{1024}$)& Dropout (0.05) \\
\hline
7  & \textbf{Activation} & ($b_s, \text{1024}$)& GELU\\
\hline
8  & \textbf{Linear}  & ($b_s, \text{512}$)& None \\
\hline
9  & \textbf{Dropout} & ($b_s, \text{512}$)& Dropout (0.1) \\
\hline
10 & \textbf{Activation} & ($b_s, \text{512}$)& GELU\\
\hline
11 & \textbf{Linear}  & ($b_s, \text{512}$)& None \\
\hline
12 & \textbf{Dropout} & ($b_s, \text{512}$)& Dropout (0.1) \\
\hline
13 & \textbf{Activation} & ($b_s, \text{450}$)& GELU\\
\hline
14 & \textbf{Linear}  & ($b_s, \text{450}$)& None \\
\hline
15 & \textbf{Dropout} & ($b_s, \text{450}$)& Dropout (0.15) \\
\hline
16 & \textbf{Activation} & ($b_s, \text{450}$)& GELU \\
\hline
17 & \textbf{Linear}  & ($b_s, \text{450}$) & None \\
\hline
\end{tabular}
\caption{Detailed architecture of the custom MLP model implemented in this work. Here, $b_s=32$ represents the batch size, $N_{ch}=16$ is the number of OFDM channels used as input to the model, $N=450$ is the number of samples in a data segment, and $K$ represents the number of OFDM data streams ($K=2$ for complex-valued OFDM data).}
\label{tab:mlp}
\end{table}

This is followed by the inverse DCT operation in order to obtain the synthetic PPG signal as follows:
\begin{equation} \label{idct}
 x[n] = \frac{1}{N} \left( X[0] + 2 \sum_{k=1}^{N-1} X[k] \cos \left( \frac{\pi}{N} \left( n + \frac{1}{2} \right) k \right) \right) \quad \\
\end{equation}
This allows us to compute the mean absolute error loss function between the synthetic PPG and reference PPG, in order to optimize the weights and biases of the MLP through backpropagation. This concludes the training process. 

Then, during the test phase, we compute the DCT type-II of an OFDM segment, pass it through MLP to get the DCT of the synthetic PPG, take the inverse DCT to obtain the digital twin PPG.

{\it Hyperparameters of the custom MLP:}
We used GELU activation function in all layers except the last layer. Further, we utilized L2 regularization with $\lambda=1e{-6}$ along with L1 (mean absolute error) loss function, in order to avoid overfitting. The optimiser used in the learning process was ADAM with a learning rate $\eta=1e{-4}$.  
Finally, we set batch size = 32 in our experiments.


\subsection{{Proposed approach: U-NET-based TD method}}

{Our primary approach is a time-domain method centered on a purely convolutional U-NET architecture. This was a deliberate design choice, tailored to the specific challenges of our radio-to-PPG translation task. The core challenge is to reconstruct a signal with a well-defined, repeating {morphological structure} of the PPG pulse. U-NET architectures are state-of-the-art for such morphological feature extraction and reconstruction tasks, proven by their success in biomedical image segmentation \cite{ronneberger2015u}. We hypothesized that the strength of convolutional kernels in learning hierarchical spatial patterns would translate directly to our 1D temporal problem, enabling the model to learn the intrinsic shape of the PPG waveform from the complex radio input. While more complex architectures like Transformers are popular, we conjectured that a U-NET provides the optimal trade-off between performance and complexity for this problem, especially given our limited-size dataset. It represents a powerful yet simpler solution, specifically engineered for structural reconstruction. Our model uses two U-NETs in cascade: an approximation network and a refinement network, a strategy effective for waveform translation tasks \cite{ibtehaz2022ppg2abp}. This two-step process first generates a coarse PPG waveform and then refines its morphological details to produce a high-fidelity output.}

\begin{figure*}[ht]
\begin{center}
	\includegraphics[width=16cm,height=5cm]{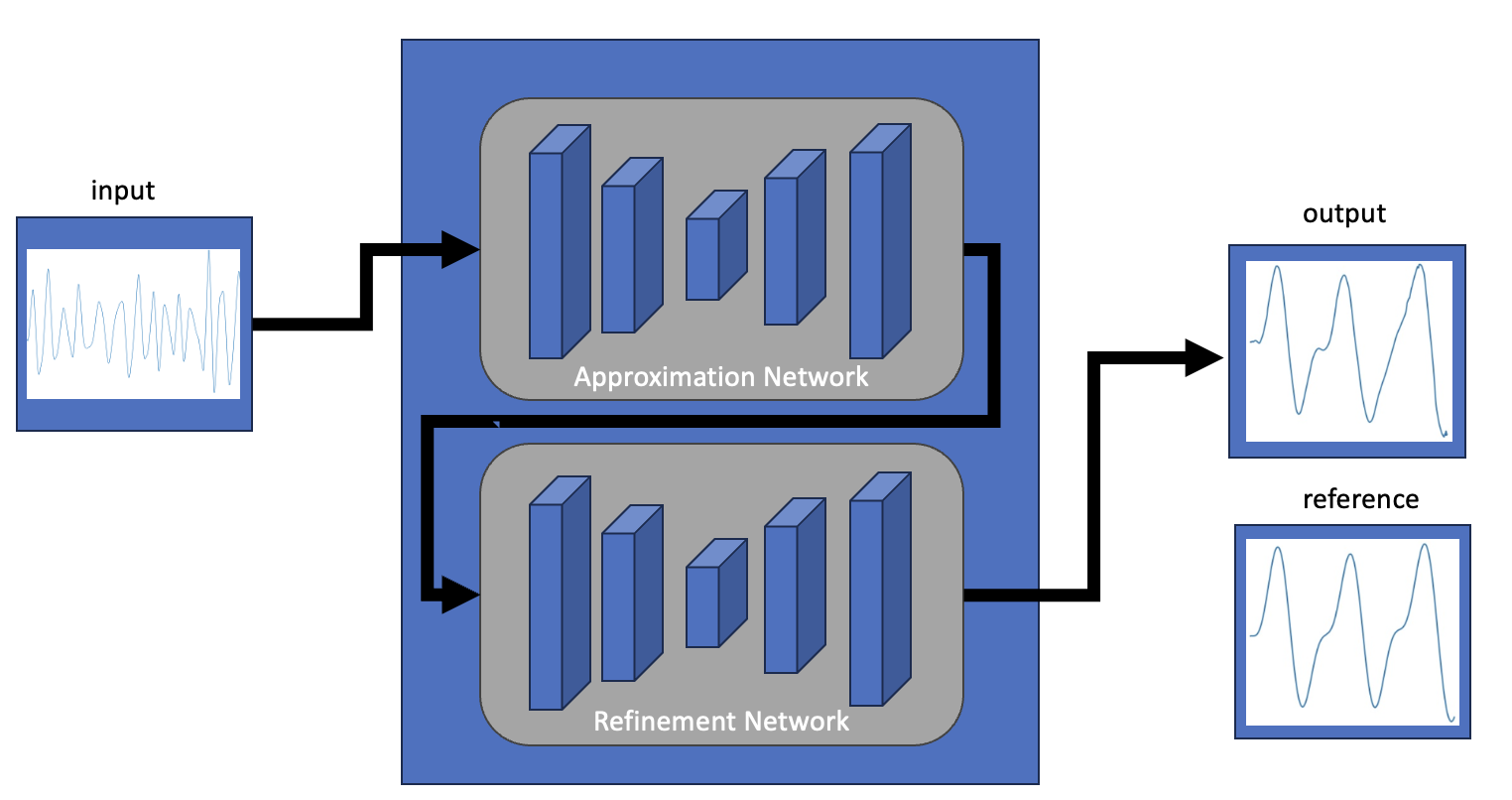} 
\caption{The custom U-NET model consists of an Approximation network and a Refinement network in cascade.}
\label{fig:UnetBlock}
\end{center}
\end{figure*}

\begin{figure}
 \centering
 \begin{subfigure}{0.85\linewidth} 
  \includegraphics[width=3.4in]{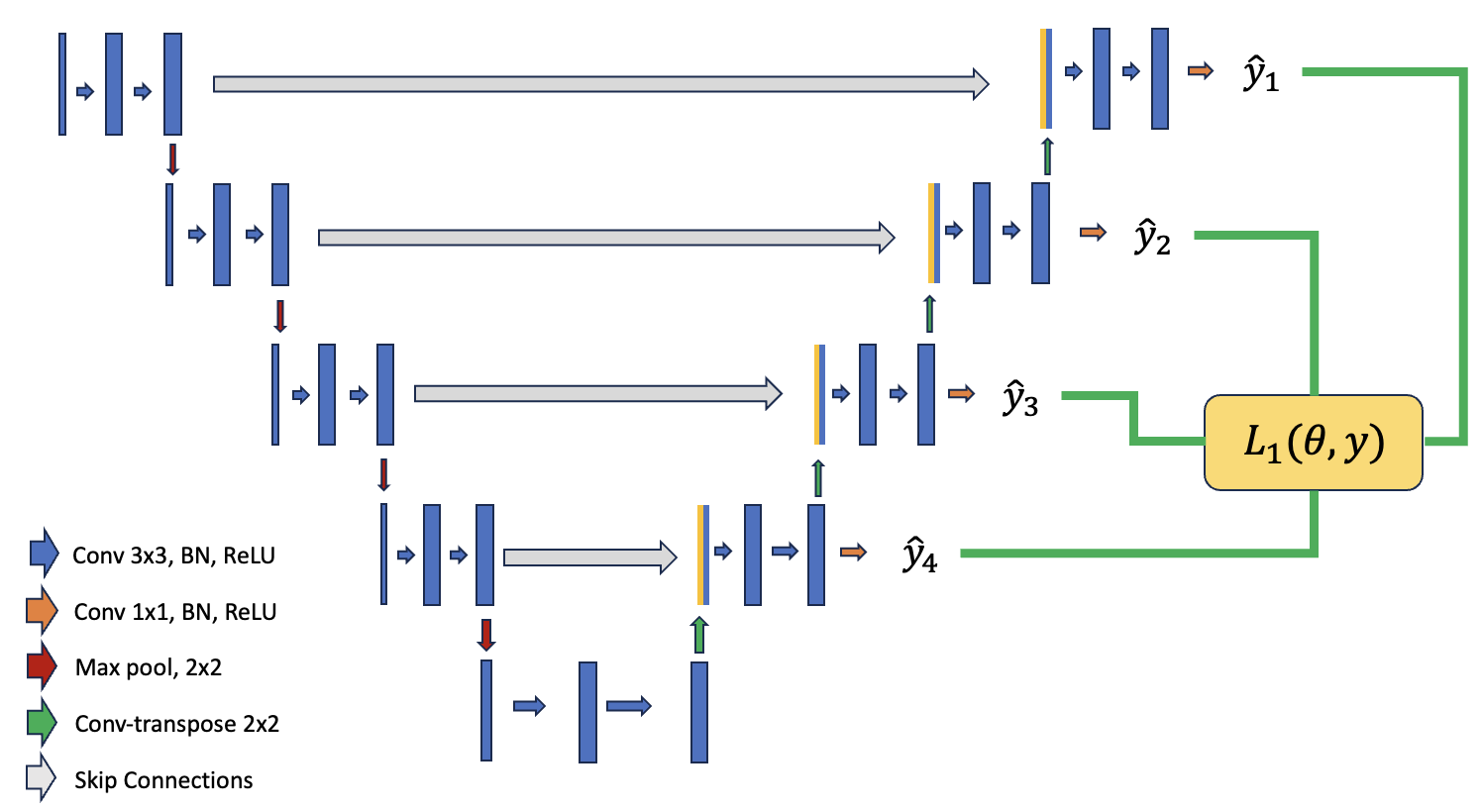}
  \caption{U-NET Approximation network.}
  \label{beforeAlign}
 \end{subfigure}
 \hfill
 \begin{subfigure}{0.85\linewidth} 
  \includegraphics[width=3.4in]{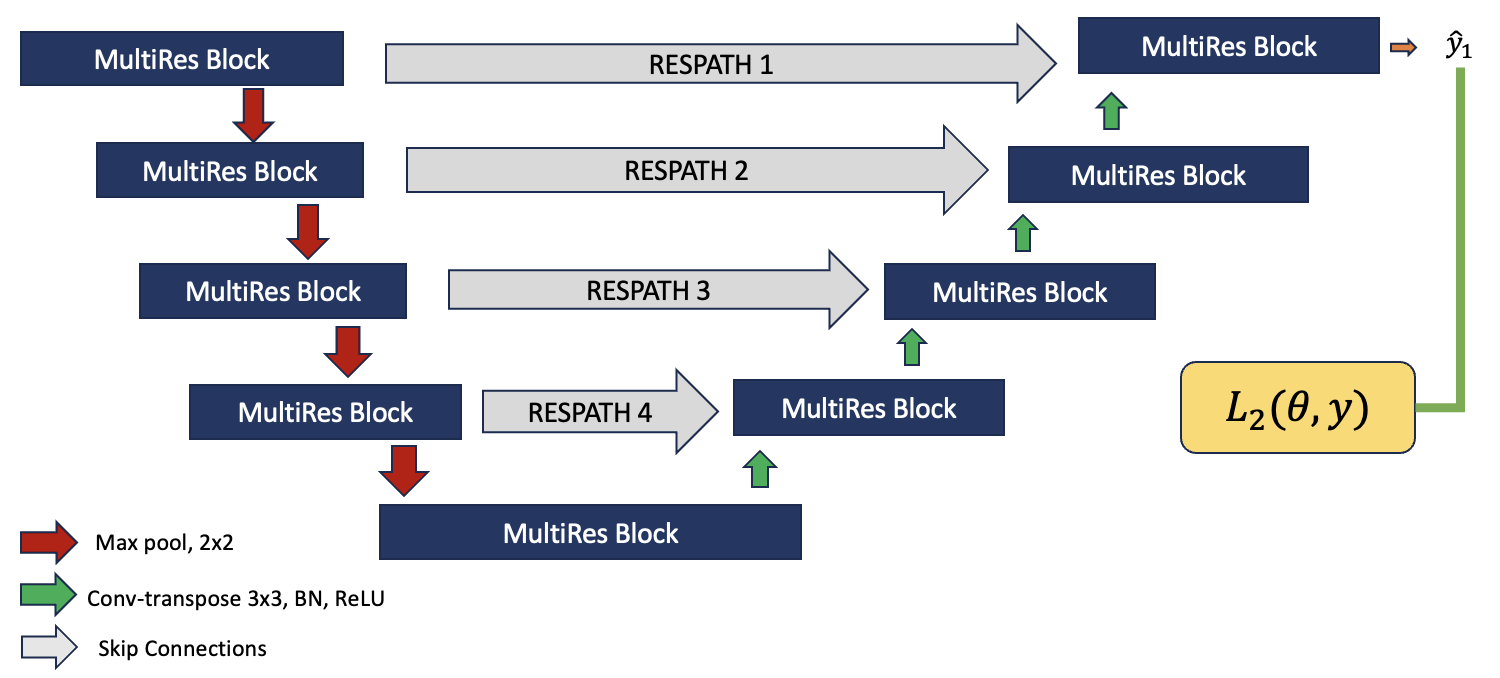}
  \caption{U-NET Refinement network.}
  \label{AfterAlign}
 \end{subfigure}
 \vfill
 \begin{subfigure}{0.85\linewidth} 
  \includegraphics[width=3.4in]{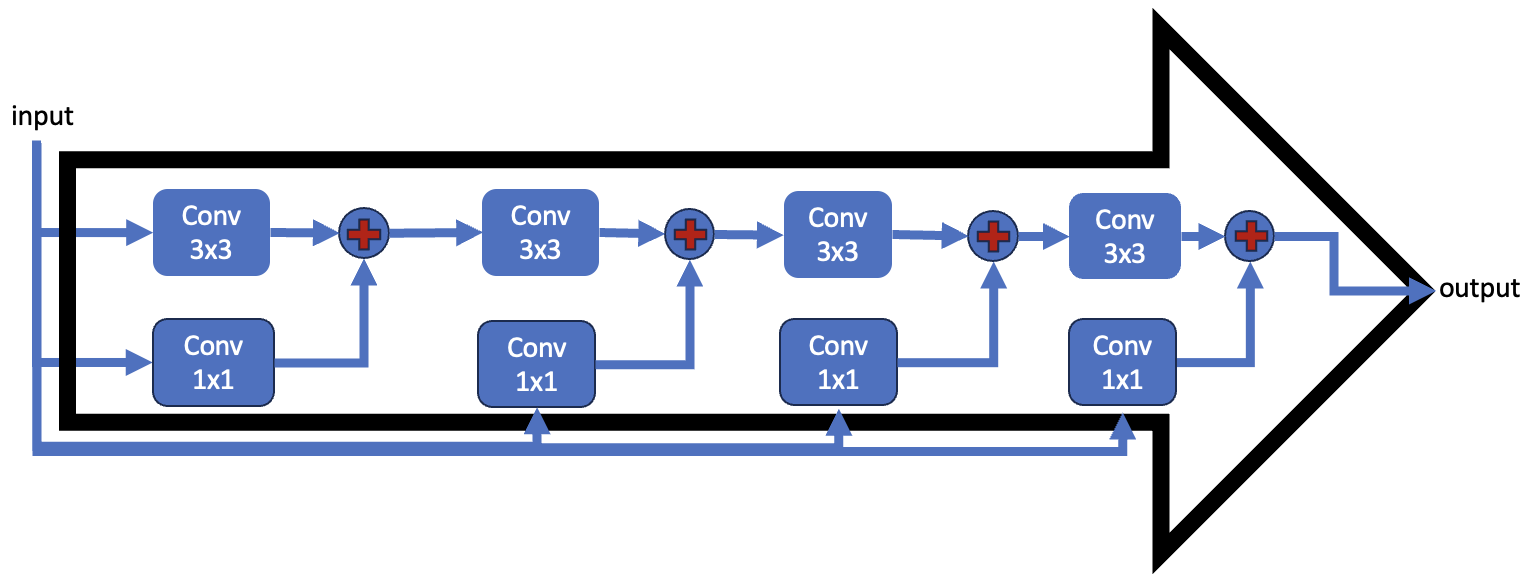}
  \caption{Respath block of Refinement network (inside view).}
  \label{AfterAlign}
 \end{subfigure}
 \hfill
 \begin{subfigure}{0.85\linewidth} 
  \includegraphics[width=3.4in]{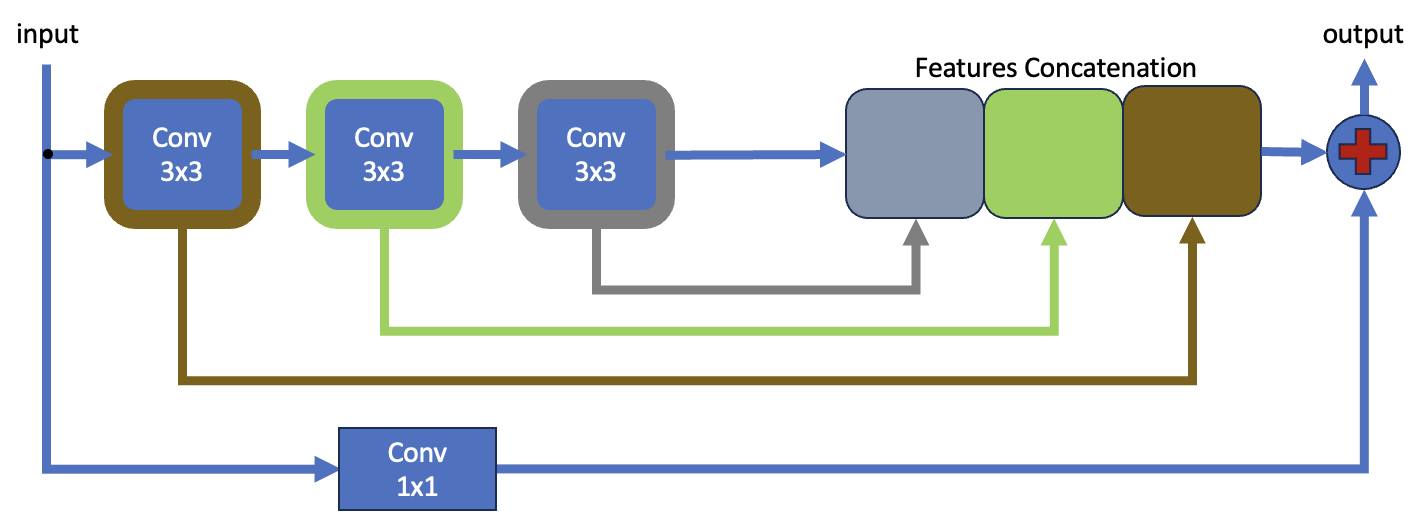}
  \caption{MultiResBlock of Refinement network (inside view).}
  \label{AfterAlign}
 \end{subfigure}
 \caption{ Zoomed-in views of the Approximation network and the Refinement network, and the sub-components. }
 \label{fig:UnetZoomIn}
\end{figure}

{Keeping in mind that we need to learn a map between two 1D signals, we repurpose the U-NET model for the 1D signals by substituting two-dimensional operations--convolution, pooling, and upsampling--with their one-dimensional equivalents. The terminal convolutional layer is modified to use a linear activation function to facilitate regression output. Furthermore, we implement deep supervision as in \cite{ibtehaz2022ppg2abp} within our U-NET configuration, which helps minimize errors by leveraging the multi-scale reconstruction errors, and enhances the training of intermediate layers by generating a subsampled version of the target signal before each up-sampling step in the decoder. These auxiliary losses play a pivotal role in refining the training output, significantly improving the accuracy of the final output.}
Nevertheless, we observe that the output of the Approximation model occasionally diverges from the reference PPG.
This motivates us to implement another model---the Refinement network---in cascade which incorporates the 1D MultiRes U-NET model (which represents an advancement over the traditional U-NET model). The primary modifications in MultiRes U-NET model include the incorporation of Multi-Residual (MultiRes) blocks and Residual (Res) paths. The MultiRes blocks are characterized by a streamlined multiresolution analysis via factorized convolutions, enhancing the network's capacity to analyze signals at multiple scales. Meanwhile, the Respaths introduce extra convolutional operations along the shortcut connections, aiming to bridge the gap between the feature maps produced by the encoder and decoder at corresponding levels. This design modification is pivotal in minimizing the disparity in feature representation across the network, thereby refining the fidelity of the reconstructed PPG signal.

{\it Design of the two custom loss functions:}
We customize the loss function $L_1$ of the Approximation network whereby we add two regularization terms into the loss function that take into account the first and second derivatives of the output signal as follows: 

\begin{equation}
\begin{split}
L_1 &= \frac{1}{N_b}\sum_{j=1}^{N_b}\left( \frac{1}{N} \sum_{i=1}^N \sum_{k=1}^4 |y_k[i] - \hat{y}_k[i]| \right. \\& + \lambda_1 \sum_{i=1}^{N-1} \sum_{k=1}^4 |y'_k[i] - \hat{y}'_k[i]| \\
&+ \left. \lambda_2 \sum_{i=1}^{N-2} \sum_{k=1}^4 |y''_k[i] - \hat{y}''_k[i]| \right)
\end{split}
\label{costFunctionUNETApp}
\end{equation}
where $\hat{y}_{i,k}$ is the output of the Approximation network at level $k$, which denotes a sub-sampled version of the reconstructed digital twin PPG signal, $N$ is the number of samples in each batch, and $N_b$ is the number of batches used to train the U-NET model. Further, $\lambda_1$, $\lambda_2$ are the regularization constants. Specifically, Eq. (\ref{costFunctionUNETApp}) consists of the following terms:  
\begin{itemize}
  \item \textit{Mean Absolute Error:} The term $\frac{1}{N} \sum_{i=1}^N \sum_{k=1}^4 |y_{k}[i] - \hat{y}_{k}[i]|$ calculates the mean absolute error between the reference PPG $y_{k}[i]$ and the output (digital twin PPG) $\hat{y}_{k}[i]$ over $N$ data points and 4 levels. This term assesses the accuracy of the model in reconstructing the digital twin PPG values across multiple levels.
  
  \item \textit{First Derivative Regularization:} The regularization term $\lambda_1 \sum_{i=1}^{N-1} \sum_{k=1}^4 |y'_{k}[i] - \hat{y}'_{k}[i]|$ adds a penalty based on the absolute differences between the first derivatives of the reference PPG and digital twin PPG signals for each of the 4 levels. This term ensures that the model captures the dynamic trends in the reference PPG at different scales.
  
  \item \textit{Second Derivative Regularization:} Similarly, $\lambda_2 \sum_{i=1}^{N-2} \sum_{k=1}^4 |y''{k}[i] - \hat{y}''{k}[i]|$ penalizes discrepancies in the second derivatives across the 4 levels, addressing the smoothness and curvature aspects of the digital twin PPG signal comprehensively.
  
  \item \textit{Batch Averaging:} The three terms in Eq. (\ref{costFunctionUNETApp}) are averaged over $N_b$ batches, indicated by $\frac{1}{N_b}\sum_{j=1}^{N_b}(...)$. This operation promotes stability in the learning process by reducing variance in the $L_A$ loss function across different data subsets.
\end{itemize}
This way, the modified cost function $L_1$ helps Approximation network reconstruct the PPG wave in a multi-level fashion, and is capable of learning time-varying features of the both signals. This in turn allows the Approximation network to keep track of the phase and amplitude variations of the reference PPG. 

Similarly, we customize the loss function $L_2$ of the Refinement network whereby we add two regularization terms into the loss function that take into account the first and second derivatives of the output signal as follows: 
\begin{equation}
\begin{split}
L_2 = \frac{1}{N_b}\sum_{j=1}^{N_b}\left( \frac{1}{N} \sum_{i=1}^N |y[i] - \hat{y}[i]| \right. &+ \lambda_1 \sum_{i=1}^{N-1} |y'[i] - \hat{y}'[i]| \\
&+ \left. \lambda_2 \sum_{i=1}^{N-2} |y''[i] - \hat{y}''[i]| \right)
\end{split}
\label{costFunctionUNETRef}
\end{equation}
where $\hat{y}$ ($y$) represents the reconstructed digital twin (reference) PPG.
The cost function $L_2$ helps the Refinement network adapt to minor morphological details of reference PPG and reflect them in the output, i.e., digital twin PPG. Further, the skip connections combined with the MultiResblock (see Fig. \ref{fig:UnetZoomIn}) help overcome the gradient vanishing problem and make the training easier. 

{\it Setting of hyperparameters $\lambda_1$ and $\lambda_2$:}
The scalars $\lambda_1$ and $\lambda_2$ are tunable hyperparameters that allow us to give more weightage to the accuracy compared to the dynamical properties of the reconstructed signal, and vice versa. In this work, both $\lambda_1$ and $\lambda_2$ are set to 1, which indicates a balanced emphasis on minimizing the absolute error in the reconstructed signal, its first derivative, and its second derivative. This setting of hyperparameters $\lambda_1$ and $\lambda_2$ ensures that our model is not only accurate in terms of signal reconstruction but also faithfully captures the dynamic properties of the PPG signal, which is something very desirable from digital twin PPG signal.

\subsection{{Validation of Digital Twin PPG}}

{We now perform a quality assessment of the digital twin PPG waveform synthesized by both the DCT+MLP and U-NET models. To evaluate performance and generalization capabilities, we employ two distinct validation schemes}: 
\begin{enumerate}
 \item \textbf{Complete Pool Validation:} {A standard evaluation using a random 80/20 train-test split of the entire dataset. This assesses the model's general performance on data with similar characteristics to the training set.}
 \item \textbf{Leave-Two-Subjects-Out (LTSO) Cross-Validation:} {A more stringent scheme where the model is trained on data from all subjects except two, which are held out entirely for testing. This process is repeated until every subject has been part of a test set, specifically validating the model's ability to generalize to unseen individuals.}
\end{enumerate}

{It is crucial to contextualize the scope of this validation. The intended application for this technology is as a \textit{non-clinical monitoring system}, analogous to modern consumer wearables. The primary goal is high-fidelity signal reconstruction, not disease detection. We reiterate that our approved ethics protocol was confined to healthy volunteers, and therefore, this study does not involve data from symptomatic patients. However, by demonstrating robust PPG synthesis, we are creating a foundational technology that holds the potential to serve as a tool for long-term monitoring and pre-assessment of certain cardiovascular conditions in the future. Our validation is conducted through two downstream tasks: i) estimation of body vitals and ii) physiological feature extraction.}

\subsubsection{Validation through Vitals Estimation}

\begin{figure}[ht]
\begin{center}
	\includegraphics[width=9cm, height=9cm]{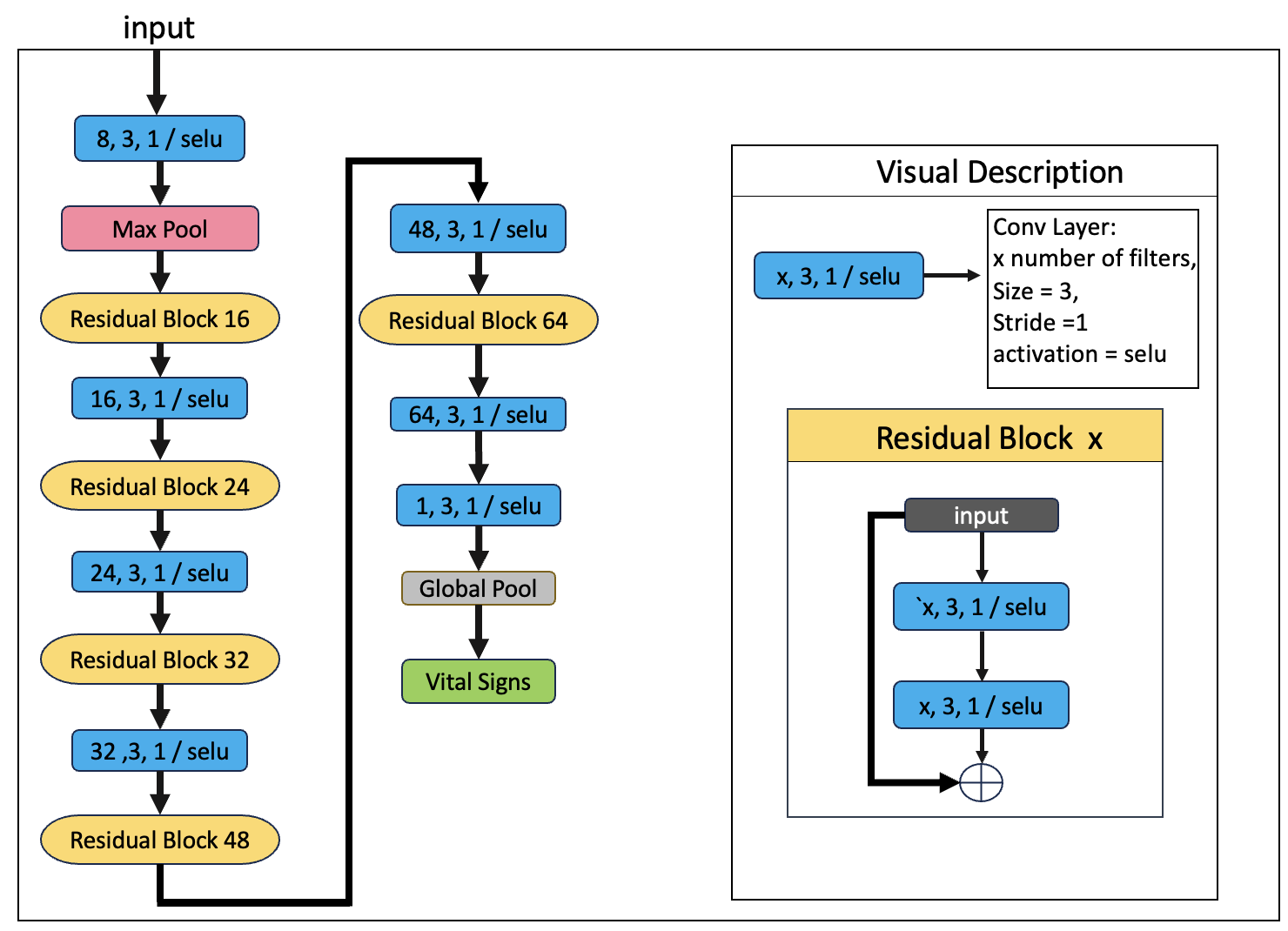} 
\caption{The deep convolutional residual neural network that we have implemented for vitals estimation.}
\label{fig:vitaModel}
\end{center}
\end{figure}

{To quantify the fidelity of the synthesized waveform, we estimate three vital signs (heart rate, breathing rate, and blood oxygen saturation) from the synthesized PPG. This is performed by feeding the synthesized signal into a baseline deep convolutional residual neural network (see Fig. \ref{fig:vitaModel}) from \cite{samavati2022efficient}. This estimation is performed on the outputs from both the \textit{Complete Pool} and the \textit{LTSO} validation schemes to provide a comprehensive view of the model's accuracy and generalization, For benchmarking, we compare the results against two baseline signals methods}:
\begin{itemize}
\item {\it Vital estimation using the raw OFDM signal:}
This method uses the same neural network presented before to estimate the set of vitals from the raw OFDM signal.
\item {\it Vitals estimation using reference PPG:}
This method uses the same neural network presented before to estimate the set of vitals from the reference PPG signal. 
\end{itemize}

{The validation stage allows us to showcase the end-to-end capabilities of the developed system whereby it collects the raw OFDM-waveform, processes it, converts it to digital twin PPG, and estimates the body vitals from it.}

\subsubsection{Validation through PPG Feature Extraction}

We also extract a number of physiological features from the digital twin PPG and its four derivative waveforms, using the open-source code \cite{cardio_repo} from github. Further, for showcasing purposes, we also extract the same set of physiological features from the reference PPG and its four derivative waveforms. The definition and detailed explanation of the features that we have extracted from the two PPG signals and their derivative signals could be found in \cite{cardio_repo}, \cite{ppg4derivatives}. 

\section{Experimental Results}

{This section presents a comprehensive performance evaluation of the two digital twin PPG synthesis methods: the frequency-domain (FD) DCT+MLP model and the time-domain (TD) U-NET model. We first analyze the synthesis quality through a multi-faceted approach: latent space visualization, and a quantitative analysis of aggregate and distributed errors. Subsequently, we assess the practical utility of the synthesized PPG via vital sign estimation.}

\subsection{Results: Digital Twin PPG Synthesis}


\subsubsection{Latent Space Visualization via t-SNE}
{To move beyond visual inspection and assess whether the synthesized signals capture the underlying structural properties of the reference PPG, we employ t-SNE. This technique visualizes high-dimensional time-series data in a low-dimensional space, revealing how well the data's intrinsic manifold is learned.}

\begin{figure}[ht!]
 \centering
 \begin{subfigure}{0.85\linewidth} 
  \includegraphics[width=2.4in]{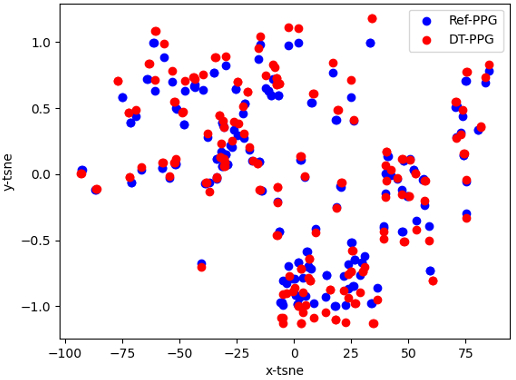}
  \caption{2D T-SNE embeddings.}
  \label{fig:2d_dct_embed}
 \end{subfigure}
 \vfill
 \begin{subfigure}{0.85\linewidth} 
  \includegraphics[width=2.4in]{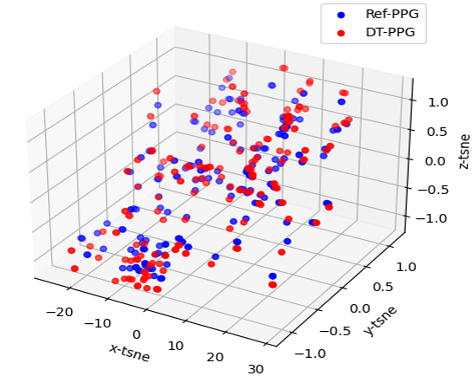}
  \caption{3D T-SNE embeddings.}
  \label{fig:3d_dct_embed}
 \end{subfigure}
 \caption{DCT+MLP method: 2D/3D T-SNE visualization of the reference PPG (blue) and digital twin PPG (red).}
 \label{fig:tsne_emb_dct}
\end{figure}

\begin{figure}[ht!]
 \centering
 \begin{subfigure}{0.85\linewidth} 
  \includegraphics[width=2.4in]{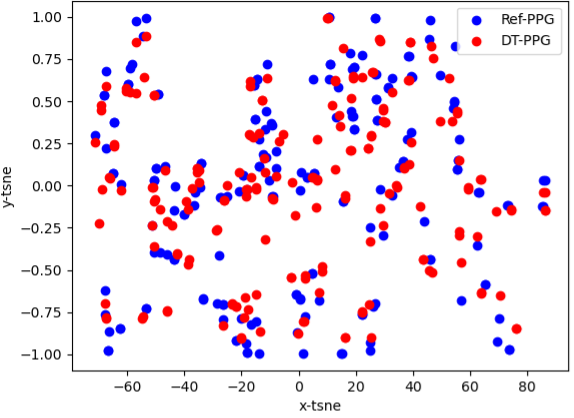}
  \caption{2D T-SNE embeddings.}
  \label{fig:2d_unet_embed}
 \end{subfigure}
 \vfill
 \begin{subfigure}{0.85\linewidth} 
  \includegraphics[width=2.4in]{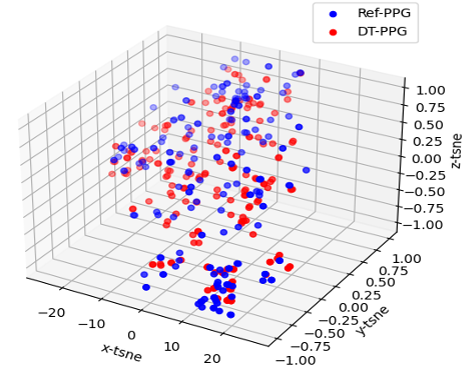}
  \caption{3D T-SNE embeddings.}
  \label{fig:3d_unet_embed}
 \end{subfigure}
 \caption{U-NET method: 2D/3D T-SNE visualization of the reference PPG (blue) and digital twin PPG (red). Note the strong intermingling of clusters, indicating high structural similarity.}
 \label{fig:tsne_emb_unet}
\end{figure}

{Figures \ref{fig:tsne_emb_dct} and \ref{fig:tsne_emb_unet} present the 2D and 3D t-SNE embeddings for the reference PPG (blue) and the synthesized digital twins from both models. For the DCT+MLP model (Fig. \ref{fig:tsne_emb_dct}), while there is proximity, some separation between the reference and synthesized clusters is visible. However, for the U-NET model (Fig. \ref{fig:tsne_emb_unet}), the embeddings from both signals are highly intermingled, forming a single, cohesive cluster. This powerful result indicates that our U-NET architecture successfully learns the complex, non-linear manifold of the PPG signal, generating a digital twin that is structurally consistent with the ground truth.}

\subsubsection{Quantitative Error Analysis}

{For a rigorous quantitative assessment, we complement aggregate error metrics with a more insightful analysis of the pointwise error distributions.}

\textit{Aggregate Metrics:} {Table \ref{tab:ppgReconstructionSummary} summarizes key error metrics (RMSE, MAE, MSE) for both models under the two validation schemes. The results clearly show the U-Net's superiority, especially in training, where its MAE is an order of magnitude lower than the DCT+MLP baseline. On the test set, particularly in the more challenging Leave-Two-Out CV scenario, the U-Net consistently maintains a lower error, demonstrating better generalization to unseen subjects.}

\begin{table*}[!ht]
\centering
\caption{{Digital twin PPG reconstruction error summary for the DCT+MLP and the final U-Net models under two validation strategies: a pooled random split and a leave-two-subjects-out cross-validation (CV). All values are rounded to four decimal places.}}
\label{tab:ppgReconstructionSummary}
\begin{tabular}{@{}llccccccc@{}}
\toprule
\multirow{2}{*}{\textbf{Split Strategy}} & \multirow{2}{*}{\textbf{Model}} & \multirow{2}{*}{\textbf{Data}} & \multirow{2}{*}{\textbf{RMSE}} & \multicolumn{3}{c}{\textbf{Mean Absolute Error (MAE)}} & \multicolumn{2}{c}{\textbf{Mean Squared Error (MSE)}} \\ 
\cmidrule(l){5-7} \cmidrule(l){8-9}
 &  & \textbf{Type} &  & \textbf{Mean} & \textbf{Std. Dev.} & \textbf{Median} & \textbf{Mean} & \textbf{Std. Dev.} \\ \midrule
\multirow{4}{*}{Pooled Random Split} & \multirow{2}{*}{DCT + MLP} & train & 0.3993 & 0.3218 & 0.2365 & 0.2762 & 0.1595 & 0.2064 \\
 &  & test & 0.6126 & 0.5081 & 0.3422 & 0.4616 & 0.3753 & 0.4362 \\ \cmidrule(l){2-9} 
 & \multirow{2}{*}{U-Net} & train & 0.0342 & 0.0250 & 0.0234 & 0.0191 & 0.0012 & 0.0032 \\
 &  & test & 0.5747 & 0.4419 & 0.3674 & 0.3435 & 0.3302 & 0.4912 \\ \midrule
\multirow{4}{*}{LTSO} & \multirow{2}{*}{DCT + MLP} & train & 0.1627 & 0.1078 & 0.1218 & 0.0705 & 0.0265 & 0.0761 \\
 &  & test & 0.6592 & 0.5220 & 0.4026 & 0.4320 & 0.4346 & 0.5977 \\ \cmidrule(l){2-9} 
 & \multirow{2}{*}{U-Net} & train & 0.0356 & 0.0261 & 0.0243 & 0.0198 & 0.0013 & 0.0032 \\
 &  & test & 0.6464 & 0.5242 & 0.3782 & 0.4545 & 0.4179 & 0.5271 \\ \bottomrule
\end{tabular}
\end{table*}

\begin{figure}[ht!]
 \centering
 \includegraphics[width=9cm, height=9cm]{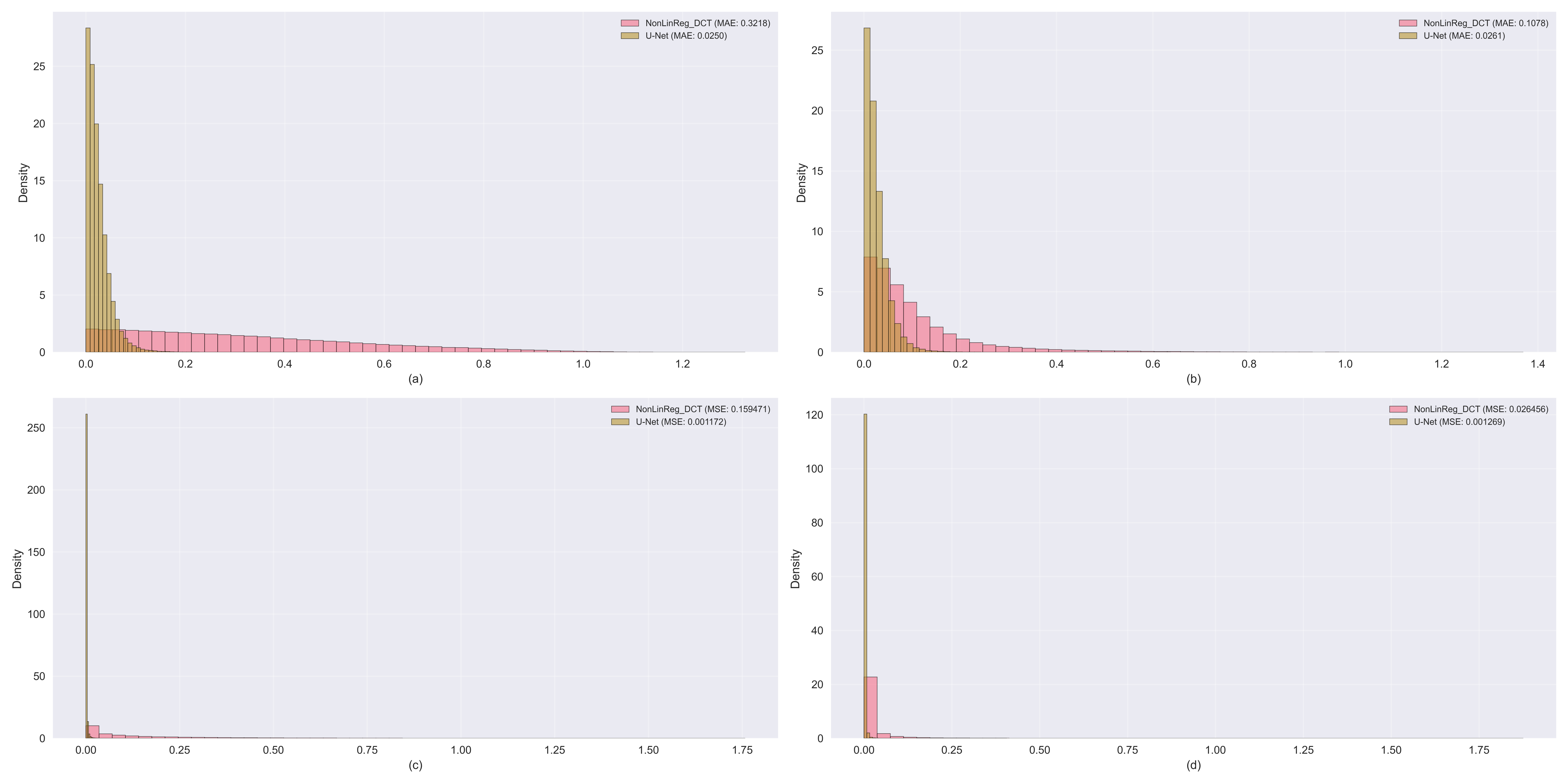}
 \caption{{Comparative histograms of pointwise reconstruction errors on the \textit{training set} for Pooled Split (a, c) and Leave-Two-Subjects-Out (b, d). The U-Net's distribution is more tightly concentrated at zero.}}
 \label{fig:train_error_hist}
\end{figure}

\begin{figure}[ht!]
 \centering
 \includegraphics[width=9cm, height=9cm]{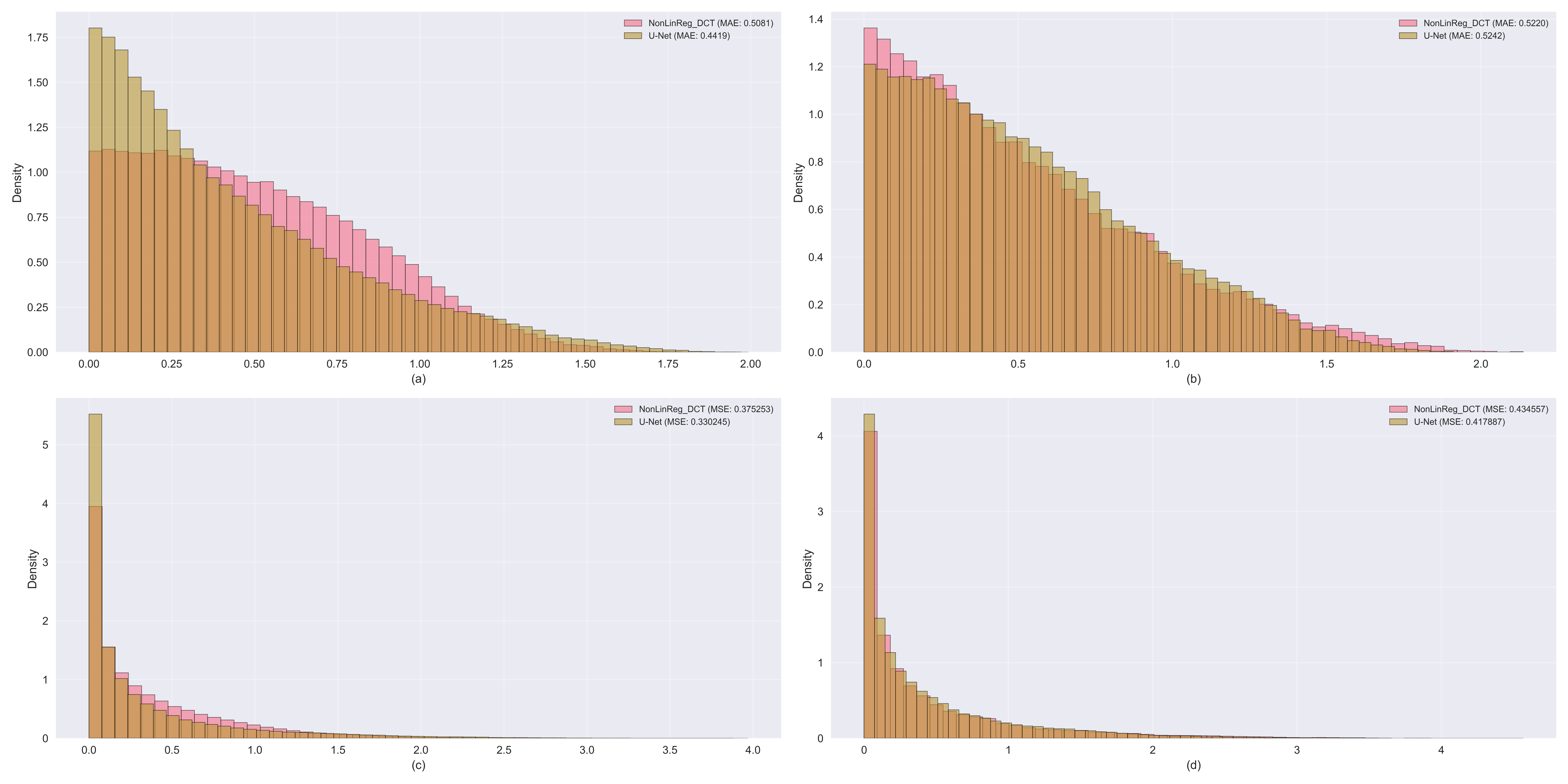}
 \caption{{Comparative histograms of pointwise reconstruction errors on the \textit{test set} for Pooled Split (a, c) and Leave-Two-Subjects-Out (b, d). The U-Net model (orange) consistently shows an error distribution more skewed towards zero, indicating superior generalization and reliability.}}
 \label{fig:test_error_hist}
\end{figure}

\textit{Pointwise Error Distributions:} {To understand model reliability beyond average performance, we analyze the distribution of pointwise errors. Figures \ref{fig:train_error_hist} and \ref{fig:test_error_hist} show these distributions for the training and test sets. While both models show low errors on training data (Fig. \ref{fig:train_error_hist}), the test data histograms (Fig. \ref{fig:test_error_hist}) are far more revealing. The DCT+MLP model exhibits a broad, heavy-tailed error distribution, indicating a frequent occurrence of significant reconstruction errors. The U-Net model, in stark contrast, maintains a distribution that is sharply concentrated near zero, with a much faster decay. This demonstrates that the U-Net is not only more accurate on average but is also significantly more reliable, making far fewer large errors. This robust quantitative evidence solidifies the superiority of our time-domain U-NET approach.}

{\it Ablation analysis on required sensing overhead:}
We develop an ablation study in which we vary the number of OFDM channels $N_{ch}$ from 1 to 16 in order to study its impact on the performance of the two deep learning models (see Table \ref{table:mae}). First of all, we observe that the U-NET based TD method registers a much lower mean absolute error (MAE) compared to DCT+MLP based FD method, for the full range of $N_{ch}$. Secondly, though the DCT+MLP method registers a minimum MAE of 2.699 for $N_{ch}=8$, it is mostly indifferent of $N_{ch}$. Thirdly, the U-NET based TD model achieves an impressive minimum MAE of \textbf{0.194} for $N_{ch}=10$ OFDM channels. We believe the superior performance of the U-NET method is due to the following reasons:
\begin{enumerate}
 \item The U-NET model utilizes both the real part and imaginary part of the $N_{ch}$-channel OFDM signal, while the DCT+MLP model only utilizes the magnitude of the $N_{ch}$-channel OFDM signal (which leads to partial loss of useful information). 
 \item The two cost functions $L_1,L_2$ utilized by the U-NET Approximation and refinement networks combine the power of multi-scale reconstruction and regularization to better learn the dynamics and curvature of the reference PPG (see Eqs. (\ref{costFunctionUNETApp}),(\ref{costFunctionUNETRef})).
\end{enumerate}

Finally, an on-demand health sensing overhead of 10 (out of 64) OFDM channels implies that 54 OFDM channels still remain available for data exchange in the 6G/WiFi ISAC context. Thus, the proposed methods for digital twin PPG synthesis incur a small ISAC overhead (15.62\%) in terms of required bandwidth, and thus, might be lucrative for Telecom operators as a new vertical/revenue stream. 

\begin{table}[ht]
\centering
\begin{tabular}{|c|c|c|}
\hline
\textbf{$N_{ch}$} & \textbf{MRAE (DCT+MLP)} & \textbf{MRAE (U-NET)} \\ \hline
2  & 2.973 $\pm{3.061}$ & 0.333 $\pm{0.196}$ \\ \hline
4  & 2.992 $\pm{2.968}$ & 0.271 $\pm{0.245}$ \\ \hline
6  & 3.001 $\pm{3.278}$ & 0.210 $\pm{0.320}$ \\ \hline
8  & {\bf 2.699} $\pm{2.962}$ & 0.202 $\pm{0.259}$ \\ \hline
10 & 3.020 $\pm{2.938}$ & {\bf 0.194 $\pm{0.149}$} \\ \hline
12 & 2.984 $\pm{2.933}$ & 0.196 $\pm{0.203}$ \\ \hline
14 & 3.103 $\pm{2.792}$ & 0.202 $\pm{0.224}$ \\ \hline
16 & 3.303 $\pm{2.978}$ & 0.197 $\pm{0.159}$ \\ \hline
\end{tabular}
\caption{Ablation study: MRAE of two AI models for test data when number of OFDM channels $N_{ch}$ is varied. }
\label{table:mae}
\end{table}

{\it Complexity analysis:}
Synthesis of high fidelity digital twin PPG comes at a cost. That is, the AI models presented in Section IV require a significant amount of computational power in order to learn complex features, such as morphology and phase of the reference PPG signal (see Table \ref{table:model_comparison}). Even though the computational cost turns out to be high, it is actually below the average for available GPU models.

\begin{table}[htbp]
\centering
\begin{tabular}{lcc}
  \toprule
  \textbf{Model} & \textbf{Trainable Parameters} & \textbf{FLOPs} \\
  \midrule
DCT+MLP & 7.69M & 184.32 MFLOPs \\
U-NET (Approximation) & 329.6K & 1.3073 GFLOPs \\
U-NET (Refinement) & 329.41K & 1.3055 GFLOPs \\
  \bottomrule
 \end{tabular}
\caption{Trainable Parameters and FLOPs (forward + backward) for the two models used for digital twin PPG synthesis. FLOPs represent the floating-point operations per second. K, M, G stand for kilo, mega, and giga, respectively. }
\label{table:model_comparison}
\end{table}

\subsection{Results: Validation of Digital Twin PPG synthesis}

{We now conduct a quality assessment of the digital twin PPG waveforms synthesized using the Discrete Cosine Transform with Multi-Layer Perceptron (DCT+MLP) based frequency domain (FD) method and the U-NET based time domain (TD) method. This assessment involves two primary experiments: i) estimation of body vitals from the digital twin PPG, and ii) feature extraction from the digital twin PPG. This section focuses on the vitals estimation.}

\subsubsection{Vitals estimation}
{To evaluate the quality of the digital twin PPG signals synthesized by the U-NET and DCT+MLP models, we estimate three vital signs: blood oxygen saturation level (SpO2), heart rate (HR), and respiratory rate (RR). These vitals are estimated by passing the synthesized PPG signals through a convolutional residual network. For benchmarking, we also estimate these vitals using two baseline methods: raw Photoplethysmography (Raw-PPG) and raw Software-Defined Radio (Raw-SDR) signals. Each method (DT-PPG U-NET, DT-PPG DCT+MLP, Raw-PPG, Raw-SDR) is evaluated under two configurations: "LTSO" and "Pool".}

We utilize the Mean Relative Absolute Error (MRAE) and Mean Relative Standard Deviation (MRSD) as performance metrics, defined as:
\begin{equation}
\text{MRAE} = \frac{1}{n} \sum_{i=1}^{n} \frac{| \text{True Value}_i - \text{Measured Value}_i |}{| \text{True Value}_i |} \times 100\%
\label{mean_rel_abs_error}
\end{equation}

\begin{equation}
\text{MRSD} = \frac{1}{n} \sum_{i=1}^{n} \frac{\text{Standard Deviation}_i}{| \text{True Value}_i |} \times 100\%
\label{mean_rel_std_error}
\end{equation}
where $n$ represents the number of examples.

{Table \ref{tab:mae-rel-table} summarizes the training and validation MRAE for all methods and configurations across the three vital signs.}

{\textit{SpO2 estimation:} The Raw-SDR (LTSO) method achieved the lowest validation MRAE of 0.00835 $\pm{0.00604}$, closely followed by Raw-SDR (Pool) at 0.00965 $\pm{0.00877}$. Among the digital twin PPG methods, DT-PPG{(U-NET)} (LTSO) showed the best performance with a validation MRAE of 0.01037 $\pm{0.00911}$, demonstrating competitive accuracy with the best baseline methods. DT-PPG{(U-NET)} (Pool) also performed well at 0.01152 $\pm{0.00908}$. In contrast, DT-PPG{(DCT+MLP)} (LTSO) exhibited a significantly higher validation MRAE of 0.04910 $\pm{0.02752}$, and Raw-PPG (Pool) showed higher MRAE of 1.11064 $\pm{0.01925}$, indicating poor performance in these specific configurations.}

{\textit{HR estimation:} The Raw-PPG (Pool) method achieved the lowest validation MRAE of 0.06430 $\pm{0.05139}$. Among the digital twin PPG methods, DT-PPG{(U-NET)} (Pool) performed best with a validation MRAE of 0.09469 $\pm{0.07636}$, which is competitive with the Raw-SDR (Pool) method (0.07368 $\pm{0.05735}$). The DT-PPG{(DCT+MLP)} (LTSO) method showed a very high validation MRAE of 1.27025 $\pm{0.02220}$, indicating a substantial error in this configuration.}

{\textit{RR estimation:} Raw-PPG (Pool) achieved the lowest validation MRAE of 0.17696 $\pm{0.14648}$. DT-PPG{(U-NET)} (Pool) demonstrated strong performance among the digital twin methods with a validation MRAE of 0.21134 $\pm{0.23166}$, closely followed by Raw-SDR (LTSO) at 0.21398 $\pm{0.16220}$ and DT-PPG{(U-NET)} (LTSO) at 0.24349 $\pm{0.19629}$. Notably, DT-PPG{(DCT+MLP)} (Pool) showed a very high validation MRAE of 5.26606 $\pm{1.22580}$, suggesting that this configuration is not suitable for accurate RR estimation.}

{Overall, the U-NET based digital twin PPG synthesis method, particularly in its "Pool" configuration, consistently demonstrates competitive performance for vital sign estimation, often approaching or matching the accuracy of the better-performing raw signal baselines. The DCT+MLP method, while showing some competitive results (e.g., for SpO2 in the Pool configuration), also exhibits instances of significantly higher errors for HR and RR estimation occasionally.}

\begin{table*}[!h]
 \centering
 \begin{tabular}{lcccccc}
 \toprule
 & \multicolumn{2}{c}{\textbf{MRAE SpO2}} & \multicolumn{2}{c}{\textbf{MRAE HR}} & \multicolumn{2}{c}{\textbf{MRAE RR}} \\
 \cmidrule(lr){2-3} \cmidrule(lr){4-5} \cmidrule(lr){6-7}
 \textbf{Input signals} & \textbf{Train} & \textbf{Valid} & \textbf{Train} & \textbf{Valid} & \textbf{Train} & \textbf{Valid} \\
 \midrule
 \textbf{DT-PPG{(U-NET)} (LTSO)} & 0.00576 $\pm{0.00518}$ & 0.01037 $\pm{0.00911}$ & 0.01978 $\pm{0.02182}$ & 0.13789 $\pm{0.07659}$ & 0.07192 $\pm{0.09452}$ & 0.24349 $\pm{0.19629}$ \\
 \textbf{DT-PPG{(U-NET)} (Pool)} & 0.00578 $\pm{0.00534}$ & 0.01152 $\pm{0.00908}$ & 0.02901 $\pm{0.03014}$ & 0.09469 $\pm{0.07636}$ & 0.10482 $\pm{0.13897}$ & 0.21134 $\pm{0.23166}$ \\
 \textbf{DT-PPG{(DCT+MLP)} (LTSO)} & 0.02862 $\pm{0.01333}$ & 0.04910 $\pm{0.02752}$ & 1.29258 $\pm{0.04405}$ & 1.27025 $\pm{0.02220}$ & 0.12158 $\pm{0.09453}$ & 0.23520 $\pm{0.18412}$ \\
 \textbf{DT-PPG{(DCT+MLP)} (Pool)} & 0.00849 $\pm{0.00710}$ & 0.01580 $\pm{0.01393}$ & 0.08960 $\pm{0.05948}$ & 0.11755 $\pm{0.06085}$ & 5.38780 $\pm{1.33929}$ & 5.26606 $\pm{1.22580}$ \\
 \textbf{Raw-PPG (LTSO)} & 0.00661 $\pm{0.00731}$ & 0.01314 $\pm{0.00769}$ & 0.10750 $\pm{0.06468}$ & 0.17739 $\pm{0.05380}$ & 0.10560 $\pm{0.09748}$ & 0.22006 $\pm{0.16891}$ \\
 \textbf{Raw-PPG (Pool)} & 1.10887 $\pm{0.02027}$ & 1.11064 $\pm{0.01925}$ & 0.05737 $\pm{0.04703}$ & 0.06430 $\pm{0.05139}$ & 0.11407 $\pm{0.11112}$ & 0.17696 $\pm{0.14648}$ \\
 \textbf{Raw-SDR (LTSO)} & 0.00957 $\pm{0.00707}$ & 0.00835 $\pm{0.00604}$ & 0.03551 $\pm{0.03392}$ & 0.12169 $\pm{0.05793}$ & 0.13648 $\pm{0.12255}$ & 0.21398 $\pm{0.16220}$ \\
 \textbf{Raw-SDR (Pool)} & 0.00859 $\pm{0.00769}$ & 0.00965 $\pm{0.00877}$ & 0.06847 $\pm{0.05683}$ & 0.07368 $\pm{0.05735}$ & 0.97201 $\pm{0.30141}$ & 0.97357 $\pm{0.27547}$ \\
 \bottomrule
 \end{tabular}
 \caption{Quality assessment of digital twin PPG: MRAE of three vitals that we have estimated from digital twin PPG (SpO2 represents blood oxygen saturation level, HR represents the heart rate, and RR represents the respiratory rate).}
 \label{tab:mae-rel-table}
\end{table*}

\begin{figure}
 \centering
 \begin{subfigure}{0.85\linewidth} 
  \includegraphics[width=2.4in]{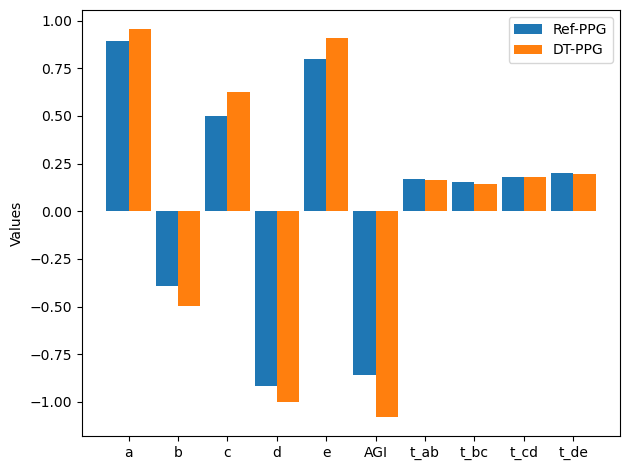}
  \caption{Feature extraction: Example 1.}
  \label{2dDctEmbed}
 \end{subfigure}
 \vfill
 \begin{subfigure}{0.85\linewidth} 
  \includegraphics[width=2.4in]{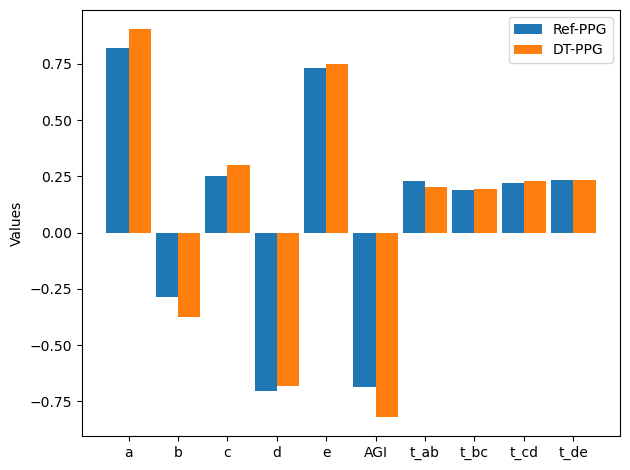}
  \caption{Feature extraction: Example 2.}
  \label{3dDctEmbed}
 \end{subfigure}
 \caption{Quality assessment of digital twin PPG: comparison of features extracted from reference PPG and digital twin PPG.}
 \label{fig:features}
\end{figure}

{\it 2. Feature extraction:}
Fig. \ref{fig:features} shows two representative examples whereby we extract a number of physiological features from the digital twin PPG as well as the reference PPG. This allows us to compare the two sets of features for the sake of quality assessment of digital twin PPG synthesized by the U-NET model. The list of extracted features in Fig. \ref{fig:features} include the set of features ''a'', ''b'', ''c'', ''d'', and ''e'', which are distinct markings on the second derivative of the PPG signal (SDPPG), and the set of features which represent the time intervals between them, i.e., $t_{ab},t_{bc},t_{cd},t_{de}$ \cite{pilt2013new}. Another interesting feature that is extracted is the aging index (AGI), calculated as: AGI = (b - c - d - e) / a. AGI is used to estimate arterial stiffness and cardiovascular aging. Fig. \ref{fig:features} demonstrates that U-NET model indeed synthesizes a high fidelity digital twin PPG (as the two sets of features are mostly in good agreement). 

\section{Discussions}

{
It is important to note that this study was conducted on healthy subjects only. While we believe the proposed framework has the potential to be extended to include pathological conditions, this would necessitate the involvement of a specialized medical team to validate the proposed AI framework for digital twin PPG synthesis, and will require a new approval from the university ethical committee for data collection from actual patients with cardiac anomalies.}

{It is generally expected that deep learning models may not perform optimally when faced with out-of-distribution data} \cite{naeini2023deep, moulaeifard2025generalizable}, {particularly with person-dependent physiological signals. The long-term performance and generalization of such systems can be significantly enhanced through fine-tuning and domain adaptation approaches, especially with the availability of larger and more diverse datasets. However, a significant challenge lies in acquiring the necessary comprehensive biomedical data for such advancements. The proposed system demonstrates real potential for deployment as a continuous health monitoring solution, having shown relative robustness in its current form.}

{This work is a significant step forward towards the design of non-contact, patient-centric health monitoring systems for smart homes and smart cities of the future. Thus, it is natural to anticipate the integration of this technology into upcoming 6G/WiFi systems. Once deployed, it will facilitate continuous, seamless, in-situ monitoring of individuals, particularly those with infectious diseases such as COVID-19, autistic people, and newborn babies.}

\section{Conclusion}

{This work introduced a novel non-contact approach for digital twin PPG signal acquisition using 6G/WiFi ISAC technology, combined with advanced deep learning models. We acquired a new dataset from 30 healthy subjects that consists of 5 hours of nearly synchronous OFDM and PPG signals, along with body vitals. This dataset was instrumental in developing and validating our AI models, i.e., the baseline DCT+MLP-based FD model and the proposed U-NET-based TD model. Among the two models, the custom-built U-NET model showed superior performance in accurately reconstructing both the morphological features and the phase of the digital twin PPG waveform. We further performed a quality assessment on the synthetic digital twin PPG signal by estimating three vital signs. It turned out that the U-NET model performed very close to the baseline methods (Raw-PPG and Raw-SDR) in terms of MRAE results, as detailed in Table \ref{tab:mae-rel-table}. It is crucial to reiterate that, to date, the Radio-PPG dataset and the proposed framework for digital twin PPG synthesis is the first of its kind.}

{Future work will focus on expanding the dataset in terms of sample size, and by incorporating data from actual patients. We will further refine the signal processing and machine learning pipeline in order to improve the robustness and generalization of the proposed system. Future work will also study the sex-related physiological differences in digital twin PPG signals, as well as the impact of sensor-to-patient distance on digital twin PPG signal quality. Last but not the least, we will extend the proposed digital twin PPG framework to other relevant downstream tasks such as identification and diagnosis of various cardiac pathologies.}

\section*{Appendix A: Setting up a USRP SDRs-based 6G/WiFi ISAC OFDM link}
\label{sec:appendixA}

The generation of OFDM signal at the transmitter involves a number of steps as follows. We first convert the serial input data stream into parallel streams. Suppose there are \( N \) channels, then we load $N$ quadrature phase shift keying (QPSK) data symbols $d_k = d[k], \quad k = 0, 1, \ldots, N-1$ onto these channels. We then apply the inverse fast Fourier transform (FFT) to the parallel data streams to transform them from the frequency domain to the time domain:
   \begin{equation}
       x[n] = \frac{1}{N} \sum_{k=0}^{N-1} d_k e^{j \frac{2\pi}{N} kn}, \quad n = 0, 1, \ldots, N-1
   \end{equation}
We then convert the parallel time-domain samples \( x[n] \) back to a serial stream for transmission. We append a cyclic prefix to the time-domain samples to mitigate inter-symbol interference:
   \begin{equation}
       x_{\text{cp}}[n] = x[n + N - L], \quad n = 0, 1, \ldots, L-1
   \end{equation}
where \( L \) is the length of the cyclic prefix\footnote{The cyclic prefix is what makes the channel matrix circulant, and thus, Eigen decomposition of the channel could be done using inverse FFT at Tx, and FFT at Rx, to mitigate the cross-talk across the channels.}. 
The transmitted OFDM signal \( s(t) \) can be represented as:
\begin{equation}
    s(t) = \sum_{n=0}^{N+L-1} x_{\text{cp}}[n] \delta(t - nT_s),
\end{equation}
where \( T_s \) is the sampling period and $\delta(t)$ is the dirac function.

At the receiver (Rx), the received signal \( r(t) \) is processed as follows: We first remove the cyclic prefix from the received samples to obtain \( y[n] \) and convert the serial stream \( y[n] \) to parallel streams. Next the FFT is applied to transform the received time-domain samples back to the frequency domain:
   \begin{equation}
       Y[k] = \sum_{n=0}^{N-1} y[n] e^{-j \frac{2\pi}{N} kn}, \quad k = 0, 1, \ldots, N-1
   \end{equation}
Finally, knowing the transmitted QPSK symbols, the channel coefficient $\hat{h}[k]$ for $k$-th channel is computed using the least squares (LS) estimation method. The LS method basically minimizes the squared error between the received signal \( y_k \) and the estimated signal \( \hat{y}_k = \hat{h}_k d_k \). The LS estimate of \( h_k \) is given by:
\begin{equation}
    \hat{h}_k = \arg \min_h \| y_k - h_k d_k \|^2, \quad k = 0, 1, \ldots, N-1
\end{equation}
For multiple observations, let \( \mathbf{y} = [y_1, y_2, \ldots, y_N]^{T} \) and \( \mathbf{x} = [x_1, x_2, \ldots, x_N]^{T} \). Then, the LS estimate is given by:
\begin{equation}
    \hat{h}_k = (\mathbf{d}_k^{H} \mathbf{d}_k)^{-1} \mathbf{d}_k^{H} \mathbf{y}_k, \quad k = 0, 1, \ldots, N-1
\end{equation}
This allows us to construct the raw channel frequency response (CFR) vector $\hat{\mathbf{h}}= [ \hat{h}_1,\hat{h}_2,\cdots, \hat{h}_N ]^H$. We use this data to train our custom-built DL models for PPG waveform synthesis.

{Table \ref{table:1} summarizes the values of the important hyper-parameters of the USRP SDR-based 6G/WiFi OFDM link that we have set up. }

\begin{table}[h!]
\centering
\begin{tabularx}{\linewidth}{|c|X|c|} 
 \hline
 {Parameter Name} & {Description} & {Value} \\
 \hline
 BOF & bits per OFDM symbol& $128$ \\
 B/sym & bits per symbol & $2$ \\
 $N$ & Number of channels & $64$ \\
 $N_D$ & Number of data channels & $52$ \\
 $N_P$ & Number of pilot channels & $12$ \\
 $N_{FFT}$ & Number of points used to compute the FFT & $64$ \\
 $N_{CP}$ & Number of samples in a cycle-prefix & $16$ \\
 $F_s$ & Sampling rate & $20,000$ samples/sec \\
 $f_c$ & Center frequency & $5.23$ GHz \\
 $IF_{Tx}$ & Interpolation factor & $250$ \\
 $DF_{Rx}$ & Decimation factor & $250$ \\
 $G_T, G_R$ & Tx/Rx gains & $40$ dB \\
 \hline
\end{tabularx}
\caption{Configuration parameters for the N210 USRP-SDR-based 6G/WiFi ISAC OFDM link used for acquisition of Radio-PPG dataset.}
\label{table:1}
\end{table}




\footnotesize{
\bibliographystyle{IEEEtran}
\bibliography{references}
}

\vfill\break

\end{document}